\documentclass[10pt,english,nofootinbib,notitlepage,superscriptaddress,aps,pra]{revtex4-1}
\usepackage[T1]{fontenc}
\usepackage[latin9]{inputenc}
\usepackage{amsmath}
\usepackage{amssymb}
\usepackage{graphicx}

\makeatletter

\@ifundefined{textcolor}{}
{%
 \definecolor{BLACK}{gray}{0}
 \definecolor{WHITE}{gray}{1}
 \definecolor{RED}{rgb}{1,0,0}
 \definecolor{GREEN}{rgb}{0,1,0}
 \definecolor{BLUE}{rgb}{0,0,1}
 \definecolor{CYAN}{cmyk}{1,0,0,0}
 \definecolor{MAGENTA}{cmyk}{0,1,0,0}
 \definecolor{YELLOW}{cmyk}{0,0,1,0}
 }

\newcommand{\ttt}{~}

\newcommand{\lqt}{$~$``}

\makeatother

\usepackage{babel}
\usepackage{amsmath}
\usepackage{centernot}

\begin{document}

\title{Pathways to relativistic curved momentum spaces: de Sitter case study}

\author{Giovanni AMELINO-CAMELIA}
\affiliation{Dipartimento di Fisica, Universit\`a di Roma {}``La Sapienza\textquotedbl{},
P.le A. Moro 2, 00185 Roma, Italy}
\affiliation{INFN, Sez.~Roma1, P.le A. Moro 2, 00185 Roma, Italy}

\author{Giulia GUBITOSI}
\affiliation{Dipartimento di Fisica, Universit\`a di Roma {}``La Sapienza\textquotedbl{},
P.le A. Moro 2, 00185 Roma, Italy}
\affiliation{INFN, Sez.~Roma1, P.le A. Moro 2, 00185 Roma, Italy}

\author{Giovanni PALMISANO}
\affiliation{Dipartimento di Fisica, Universit\`a di Roma {}``La Sapienza\textquotedbl{},
P.le A. Moro 2, 00185 Roma, Italy}
\affiliation{INFN, Sez.~Roma1, P.le A. Moro 2, 00185 Roma, Italy}

\begin{abstract}
Several arguments suggest that the Planck scale could be the characteristic
scale of curvature of momentum space. As other recent studies we assume
that the metric of momentum space determines the condition of on-shellness while
the momentum-space affine connection
governs the form of the law of composition of momenta.
We show that the possible choices of laws of composition of momenta are more numerous than
the possible choices of affine connection on a momentum space.
This motivates us to propose a new prescription for associating an affine connection to momentum composition, which
we compare to the one most used in the recent literature.
We find that the two prescriptions lead to the same picture of the so-called
$\kappa$-momentum space, with de Sitter metric and $\kappa$-Poincar\'e connection.
We also examine in greater detail than ever before the DSR-relativistic properties of $\kappa$-momentum space,
particularly in relation to its noncommutative law of composition of momenta.
We then show that in the case of ``proper de Sitter momentum space", with the de Sitter metric and its
Levi-Civita connection, the two prescriptions are inequivalent. Our novel prescription leads to a picture
of proper de Sitter momentum space which is DSR-relativistic and is characterized by a commutative
law of composition of momenta, a possibility for which
no explicit curved-momentum-space picture had been previously found. 
We argue that our construction provides a natural test case for the study
of momentum spaces with commutative, and yet deformed, laws of composition of momenta. Moreover, it can
serve as laboratory for the exploration of the properties of DSR-relativistic theories which are
not connected to group-manifold momentum spaces and Hopf algebras.\\

\end{abstract}
\maketitle
\tableofcontents{}

\section{Introduction}
Max Born argued in  1938~\cite{born1938}, inspired by Born duality, that curvature of momentum space
might be a needed step toward quantum gravity.
For several decades this proposal attracted little or no interest (see, however, Ref.~\cite{golfand}),
but over the last decade several independent arguments pointed more or less explicitly toward a role for
the Planck scale in characterizing a non-trivial geometry of momentum space (see, {\it e.g.}, Refs.~\cite{majidCURVATURE,dsr1Edsr2,jurekDSMOMENTUM,girelliCURVATURE,schullerCURVATURE,changMINIC,principle,grf2nd}).
Among the reasons of interest in this possibility we should mention
 approaches to the study of the quantum-gravity problem based on
 spacetime noncommutativity, particularly when considering models with ``Lie-algebra spacetime noncommutativity", $[x_\mu , x_\nu]= i \zeta^\sigma_{\mu \nu} x_\sigma$, where
 the momentum space on which spacetime coordinates generate translations is evidently curved (see, {\it e.g.}, Ref~\cite{gacmaj}).
 Also in the Loop Quantum Gravity approach~\cite{rovelliLRR} one can adopt a perspective suggesting momentum-space curvature (see, {\it e.g.}, Ref~\cite{leeCURVEDMOMENTUM}).
 And one should take notice of the fact that the only quantum gravity we actually know
 how to solve,
quantum gravity in the 2+1-dimensional case, definitely does predict a curved momentum space (see, {\it e.g.}, Refs.\ttt\cite{matschull1,matschull2,dsr3dFREIDLIVINE,stefanoCQG2013}).

We here focus on the perspective on Planck-scale-curved momentum spaces
adopted in the recently proposed \lqt relative locality framework"\ttt\cite{principle}, which
essentially abstracts the insight gained in the study of 2+1D quantum gravity,
providing a picture for how the geometry of momentum space can play a role in describing
Planck-scale-deformed relativistic kinematics.
 This proposal links the metric
on momentum space to the form of the on-shell/dispersion relation,
while the affine connection on momentum space
is linked to the form of the law of composition of momenta, which in turn determines the energy-momentum
conservation laws.

One of the issues that is most relevant for the analysis we here
report concerns~\cite{GACarXiv11105081,MercatiCarmonaCortes}
 the identification of the requirements that must be enforced on the geometry of momentum
space in order to allow the formulation of relativistic theories.
 Since special-relativistic laws of transformation cannot be symmetries
of any curved momentum space,  relativistic
invariance must be inevitably implemented according to the proposal
of {}``DSR relativistic theories\textquotedbl{}~\cite{dsr1Edsr2}
(also see Refs.~\cite{jurekDSRfirst,leejoaoPRDdsr,gacDSRnature,leejoaoCQGrainbow,jurekDSRreview,gacSYMMETRYreview}),
theories with two relativistic invariants, the speed-of-light scale
$c$ and a length/inverse-momentum scale: the scale that characterizes
the geometry of momentum space must in fact be an invariant if the
theories on such momentum spaces are to be relativistic.
Several grey areas however remain toward the understanding
of the compatibility between metric and affine connection on momentum space
that must be enforced in order to have a relativistic picture.

The other aspect which is of strong interest to us is the link between
 affine connection on momentum space
and law of composition of momenta. We here show
that the possible choices of laws of composition of momenta are more numerous than
the possible choices of affine connection on a momentum space, an issue which was not
noticed in the previous related literature and which we feel should play an important role in future
work in this research area.
Partly inspired by this observation, we here
propose a new prescription for associating an affine connection to momentum composition, which
we compare to the one most used in the recent literature.
As arena for comparing the two prescriptions we focus on the case of momentum spaces
with de Sitter metric.
We find that the two prescriptions lead to the same picture of the so-called
$\kappa$-momentum space, with de Sitter metric and $\kappa$-Poincar\'e connection.
We also examine in greater detail than ever before the DSR-relativistic properties of $\kappa$-momentum space,
particularly in relation to its noncommutative law of composition of momenta.
We then show that in the case of ``proper de Sitter momentum space", with the de Sitter metric and its
Levi-Civita connection, the two prescriptions are inequivalent. Our novel prescription leads to a picture
of proper de Sitter momentum space which is DSR-relativistic and is characterized by a commutative
law of composition of momenta.

As it will become clearer as we go along,
one of the elements of interest that motivated our analysis is the search of a natural
candidate of momentum space with a commutative composition law.
All the momentum spaces that have attracted attention so far have noncommutative composition law,
but it is an interesting hypothesis for physics the one in which the composition law is indeed
deformed but still preserves the property of commutativity.
We argue that our construction of a ``proper de Sitter momentum space" provides the first natural
example
of momentum space with commutative, and yet deformed, law of composition of momenta.

 In order to keep our presentation clear and compact we focus on the 1+1-dimensional case,
 where all the key conceptual challenges are already present
 but formulas are more compact and derivations are less tedious.

The next section sets up the analysis by reviewing the previously most studied prescription
and characterizing our novel prescription for associating a geometry of momentum space to
on-shellness and momentum composition.

In Section III we discuss the differences between these two prescriptions and we
establish the fact that
the possible choices of laws of composition of momenta are more numerous than
the possible choices of affine connection on a momentum space.

Then in Section IV we re-derive the structure of momentum space which can be inspired
by the structure of the $\kappa$-Poincar\'e Hopf algebra. This leads to the already much studied $\kappa$-momentum
space, with de Sitter metric and a torsionful affine connection. We also show that actually
the prescriptions so far given in formulating the geometry of momentum space within the
relative-locality framework are affected by an ambiguity. This ambiguity is merely academic since
the alternative geometries allowed by it give rise to equivalent relativistic kinematics,
but for momentum spaces with torsion it weakens the link from the observables of the theory
to the geometry of momentum space. We discuss how, exploiting this ambiguity, one can find two alternative
formulations of the $\kappa$-momentum space, both with the same momentum space metric and with the same
physical predictions, but formulated in terms of different affine connections.

Section V proposes our new \lqt proper de Sitter
momentum space", obtained, in the sense of our novel prescription, by adopting the de Sitter
metric in combination with its Levi-Civita connection. For this momentum space, which had not been studied
before, we derive several results which should be valuable for future studies.
Still in Section V we derive the law of composition of momenta that would
be attributed to ``proper de Sitter
momentum space" if adopting the prescription alternative to the one we are
here advocating.

Some of the main results of our analysis are located in Section VI. There we establish and we characterize
the (DSR-)relativistic properties of the novel proper de Sitter momentum space. And also for the
already well known $\kappa$-momentum
space we analyze relativistic properties in greater depth than ever before, establishing more firmly
its DSR-relativistic compatibility but also highlighting more vividly the peculiarities of
the associated relativistic theory.

While most of our results are obtained adopting a choice of coordinatization of momentum space,
the findings about DSR-relativistic compatibility of our de Sitter momentum spaces are independent of such
choices. This is shown in Section VII. We do not settle the issue of whether or not
 theories on curved momentum space are (or should be) diffeomorphism invariant, but we find that
specifically the property of momentum spaces of being relativistically compatible ({\it i.e.} compatible
with the formulation of DSR-relativistic theories on that momentum space) is a \lqt geometric property",
indeed diffeomorphism invariant.
Section VIII summarizes our findings and offers some expectations
for the possible development of this research area.

\section{Geometry of momentum space from on-shell relation and momentum conservation}
\label{sec:geometry}
The conceptual challenge which is at center stage in our study is the one of providing
a suitable geometrical interpretation of relativistic kinematics. Of course, this is not a particularly
interesting challenge in the presently adopted description of relativistic kinematics, for which it is
evidently appropriate to adopt an interpretation based on a Minkowskian momentum space:
 \begin{equation}
\begin{cases}
m^{2}=p_{0}^{2}-|\vec{p}|^{2}=\eta^{\mu\nu}p_{\mu}p_{\nu}\\
p=q+k\end{cases}\label{eq:The Laws (flat)}\end{equation}
where, for definiteness, we specialized the conservation law to the
case of a three-particle event (a two-body decay).

 But it does turn into
a highly nontrivial challenge when contemplating, as done in part of the quantum-gravity literature,
modifications of special-relativistic kinematics, with the on-shellness
and the law of energy-momentum conservation taking in general the form
\begin{equation}
\begin{cases}
m^{2}=d_{\ell}^{2}(p)\\
p=q\oplus_{\ell}k\end{cases}\label{eq:The Laws (def)}\end{equation}
 where $d_{\ell}^{2}$ and $\oplus_{\ell}$ are
functions of the components of the involved momenta and of the scale $\ell$, here assumed
to be the inverse of the quantum-gravity scale (the dependence of  $\oplus_{\ell}$ on $q$ and $k$
could be rendered more explicit in the notation by writing   $p=q\, \oplus_{\ell}\!\!(q,k) \,k$ but we opt for leaving
it implicit in order to keep our notation agile).

\subsection{Geometrical Interpretation of the on-shell relation}\label{jocnn}
The conceptual perspective of the relative-locality framework \cite{principle,grf2nd} provides
an interpretation based on the geometry of momentum space for deformations of on-shellness and of momentum-conservation
laws. The metric
$g^{\mu\nu}$ on momentum space is linked to the on-shell relation
while the affine connection on momentum space $\Gamma^{\lambda\mu}{}_{\nu}$
is linked to the law of composition of momenta $\oplus_{\ell}$, which
is the core ingredient of laws of conservation of energy-momentum.

According to Ref. \cite{principle} the link between on-shellness
and metric on momentum space is to be established by describing $d_{\ell}^{2}(p)$
as distance of $p$ from the origin of momentum space, distance given
in terms of the momentum-space metric. In formulas this means that
\begin{equation}
m^{2}=d_{\ell}^{2}\left(p,0\right)=\int dt\sqrt{g^{\mu\nu}(\gamma^{[A;p]}(t))\dot{\gamma}_{\mu}^{[A;p]}(t)\dot{\gamma}_{\nu}^{[A;p]}(t)}\label{eq: On-shell is the length}\end{equation}
 where $g^{\mu\nu}$ is the momentum-space metric and $\gamma^{[A;p]}(t)$
is the metric geodesics connecting the point $p$ to the origin
of momentum space. For the metric geodesics one has that \begin{equation}
\frac{d^{2}\gamma_{\lambda}^{[A]}(t)}{dt^{2}}+A^{\mu\nu}{}_{\lambda}\frac{d\gamma_{\mu}^{[A]}(t)}{dt}\frac{d\gamma_{\nu}^{[A]}(t)}{dt}=0\label{levicivijoc}\end{equation}
 where $A^{\mu\nu}{}_{\lambda}$ is the Levi-Civita connection.

Note that in general the relative-locality framework allows (also see
the next subsection) for the adoption of an affine connection which may
not be the Levi-Civita connection. Even in such cases the Levi-Civita
connection $A^{\mu\nu}{}_{\lambda}$ governs (\ref{levicivijoc})
(and therefore governs the on-shellness), while the affine connection
$\Gamma^{\mu\nu}{}_{\lambda}$ governs the law of composition of momenta.

In addition to the metric geodesics (\ref{levicivijoc}), which evidently
play a pivotal role in the relative-locality framework, we shall here
also consider the possible role of connection geodesics \begin{equation}
\frac{d^{2}\gamma_{\lambda}(t)}{dt^{2}}+\Gamma^{\mu\nu}{}_{\lambda}\frac{d\gamma_{\mu}(t)}{dt}\frac{d\gamma_{\nu}(t)}{dt}=0 \, .
\label{eq:Connection Geodesic}\end{equation}

\subsection{Standard geometrical interpretation of the Composition Law}\label{sec:standard Geometrical interpretation of the Composition Law}
Ref. \cite{principle} also introduces an affine connection on momentum
space, $\Gamma^{\mu\nu}{}_{\lambda}$, through the following definition
\begin{equation}
\Gamma^{\mu\nu}{}_{\lambda}(p)=-\frac{\partial}{\partial q_{\mu}}\frac{\partial}{\partial k_{\nu}}\left(q\oplus_{\ell}^{[p]}k\right)_{\lambda}|_{q=k=p}\label{eq: Connection Introduction}\end{equation}
 where \begin{equation}
q\oplus_{\ell}^{[p]}k=p\oplus_{\ell}\left[\left(\ominus_{\ell}\,p\oplus_{\ell}q\right)\oplus_{\ell}\left(\ominus_{\ell}\,p\oplus_{\ell}k\right)\right]\label{affinedef}\end{equation}
 and $\ominus_{\ell}$ is the so-called {}``antipode\textquotedbl{}
operation of $\oplus_{\ell}$, such that%
\footnote{Notice that in general the momentum-space affine connection may in
some cases be such that $p\oplus_{\ell}k\neq k\oplus_{\ell}p$, but
even in such cases one has that when $p\oplus_{\ell}k=0$ then also
$k\oplus_{\ell}p=0$.
} \begin{equation}
\left(\ominus_{\ell}p\right)\oplus_{\ell}p=0=p\oplus_{\ell}\left(\ominus_{\ell}p\right)\end{equation}

As observed in Ref.\cite{principle}, at least for the leading-order-in-$\ell$ approximation 
 of (\ref{eq: Connection Introduction}), the composition rule $q\oplus_{\ell} k$ can be interpreted 
 in terms of the parallel transport of $k$ along the geodesic connecting the origin of momentum space to $q$, {\it i.e.}
\begin{equation}
(q\oplus_{\ell} k) _{\lambda} \simeq q_{\lambda} +k_{\alpha}\tau^{\alpha}_{\lambda}(q)
\end{equation}
where $\tau$ is the parallel transport operator, whose first-order expression is:
\begin{equation}
\tau^{\alpha}_{\lambda}(q)=\delta^{\alpha}_{\lambda}-\Gamma^{\beta\alpha}\,_{\lambda}
\end{equation}
Note that this parallel transport is consistent with the following definition of the covariant derivative of a vector $V_{\alpha}$:
\begin{equation}
\nabla^{\lambda}V_{\alpha}=\partial^{\lambda}V_{\alpha}+\Gamma^{\lambda\beta}\,_{\alpha}V_{\beta}
\end{equation}

\subsection{A novel geometrical interpretation of the Composition Law}\label{sec:newGeometrical interpretation of the Composition Law}
The geometrical interpretations of on-shellness and composition laws reviewed in the previous two subsections
have already been adopted in several studies, and we shall refer to them as aspects of the ``standard geometrical interpretation".
For reasons that will become clearer as we go along we are going to consider here also an alternative geometrical interpretation.
This novel geometrical interpretation still links the metric of momentum space and the on-shellness
as described in Subsection \ref{jocnn}, but, building on a proposal first put forward preliminarily
by Mercati \cite{flaviotalk},
it adopts a description of the link between composition law and
affine connection which is in general different from the one suggested
by (\ref{eq: Connection Introduction}).

Our perspective on the geometrical interpretation of the composition law associates to
the points $q$ and $k$ the connection geodesics $\gamma^{(q)}$
and $\gamma^{(k)}$ which connect them to the origin of momentum space.
Then one introduces also a third curve $\bar{\gamma}(s)$, which we
call the parallel transport of $\gamma^{(k)}(s)$ along $\gamma^{(q)}(t)$,
such that for any given value $\bar{s}$ of the parameter $s$ one
has that the tangent vector $\frac{d}{ds}\bar{\gamma}(\bar{s})$ is
the parallel transport of the tangent vector $\frac{d}{ds}\gamma^{(k)}(\bar{s})$
along the geodesic connecting $\gamma^{(k)}(\bar{s})$ to $\bar{\gamma}(\bar{s})$.
Then the composition law is defined as the extremal point of $\bar{\gamma}$,
that is: \[
q\oplus_{\ell}k=\bar{\gamma}(1)\]
 We also illustrate this prescription in Fig.1.

\begin{figure}[htbp]
\includegraphics[scale=0.5]{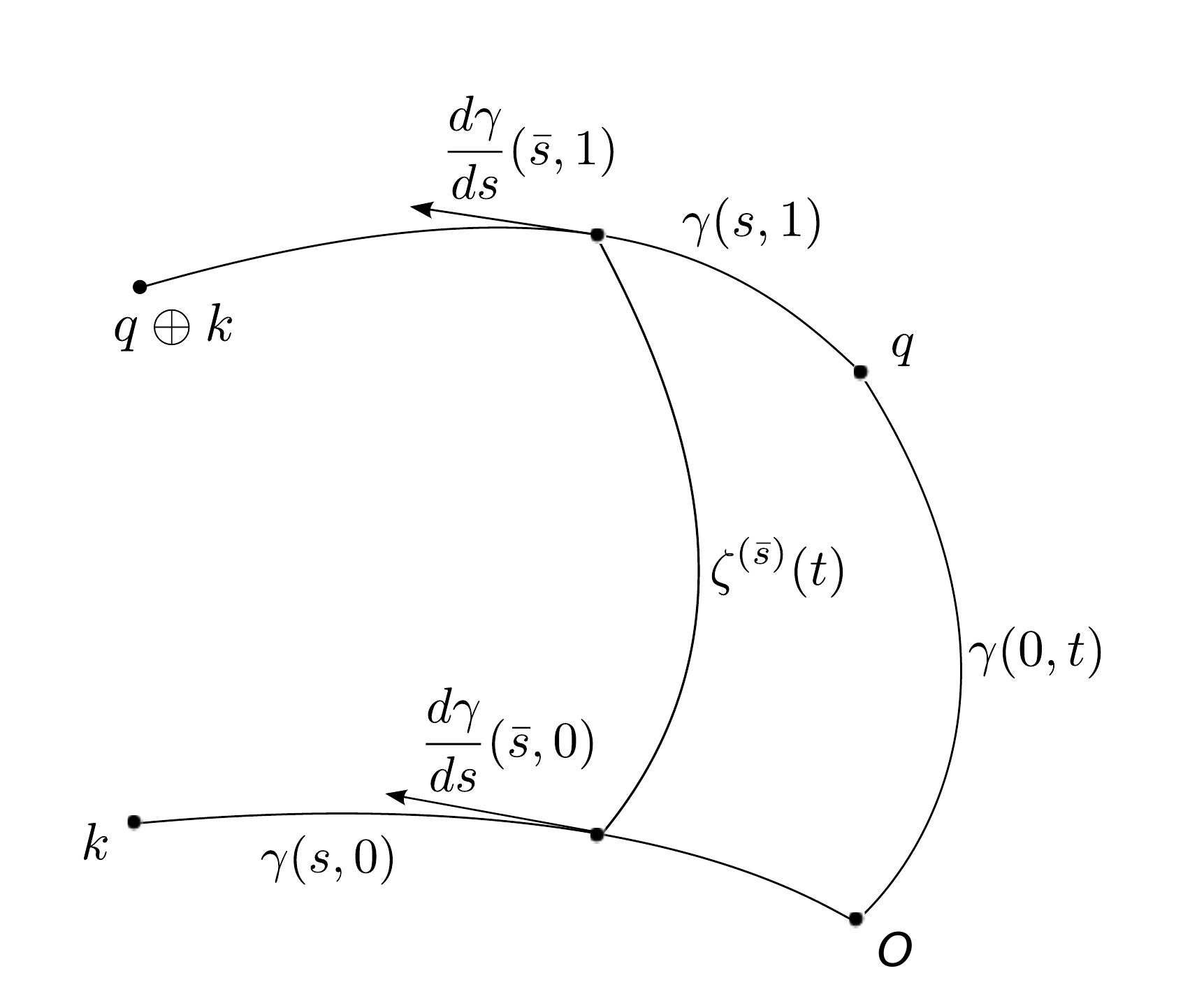}
\caption{
We determine the law of composition of momenta from the
affine connection by  associating to
the points $q$ and $k$ of momentum space the connection geodesics $\gamma^{(q)}$
and $\gamma^{(k)}$ which connect them to the origin of momentum space.
We then introduce a third curve $\bar{\gamma}(s)$, which we
call the parallel transport of $\gamma^{(k)}(s)$ along $\gamma^{(q)}(t)$,
such that for any given value $\bar{s}$ of the parameter $s$ one
has that the tangent vector $\frac{d}{ds}\bar{\gamma}(\bar{s})$ is
the parallel transport of the tangent vector $\frac{d}{ds}\gamma^{(k)}(\bar{s})$
along the geodesic connecting $\gamma^{(k)}(\bar{s})$ to $\bar{\gamma}(\bar{s})$.
Then the composition law is defined as the extremal point of $\bar{\gamma}$,
that is $q\oplus_{\ell}k=\bar{\gamma}(1)$. }
\end{figure}

A useful tool for setting up a computational approach based on this
perspective is the introduction of a parametrized surface $\gamma(s,t)$
which is related to $\bar{\gamma}$ by $\gamma(s,1)=\bar{\gamma}(s)$
and is constrained by the following boundary conditions: \begin{equation}
\begin{cases}
\gamma(s,0)=\gamma^{(k)}(s)\\
\gamma(0,t)=\gamma^{(q)}(t)\end{cases}\label{eq: comp boundary conditions}\end{equation}

The condition on the tangent vector of $\bar{\gamma}$ provides then
a condition for $\gamma(s,t)$: we want the vector $\frac{d\gamma}{ds}(s,t)$
to be the parallel transport of the vector $\frac{d\gamma}{ds}(s,0)$
along the geodesic $\zeta^{(s)}$, defined for any $s$, connecting
$\gamma(s,0)$ to $\gamma(s,1)$. This gives
\begin{equation}
\frac{d}{dt}\frac{d}{ds}\gamma_{\lambda}(s,t)+\Gamma^{\mu\nu}{}_{\lambda}(\zeta_{s}(t))\frac{d\zeta_{\mu}^{(s)}(t)}{dt}\frac{d\gamma_{\nu}(s,t)}{ds}=0\label{eq: comp from conn}\end{equation}
 where $\zeta^{(s)}(t)$ is the geodesic connecting $\gamma(s,0)$
to $\gamma(s,1)$, \textit{i.e.} \begin{equation}
\begin{cases}
\frac{d^{2}}{dt^{2}}\zeta_{\lambda}^{(s)}(t)+\Gamma^{\mu\nu}{}_{\lambda}(\zeta^{(s)}(t))\frac{d\zeta_{\mu}^{(s)}(t)}{dt}\frac{d\zeta_{\nu}^{(s)}(t)}{dt}=0\\
\zeta^{(s)}(0)=\gamma^{(k)}(s)\\
\zeta^{(s)}(1)=\gamma(s,1)\end{cases}\label{eq: intergeodesics}\end{equation}

The composition law is then defined as: \begin{equation}
q\oplus_{\ell}k=\gamma(1,1)\label{eq: comp is the extr}\end{equation}

We note that equation (\ref{eq: comp from conn}) can also be written
(in an equivalent but more explicitly covariant way) as \begin{equation}
\frac{d\zeta_{\mu}(s,t)}{dt}\nabla_{\Gamma}^{\mu}\frac{d\gamma_{\lambda}(s,t)}{ds}=0\label{eq:comp from conn cov}\end{equation}
 where $\nabla_{\Gamma}$ is the covariant derivative associated to
the connection $\Gamma$. In this form it is made explicit the fact
that the covariant derivative of the vector $\frac{d\gamma}{ds}(s,t)$
vanishes along the direction of the tangent vector $\frac{d\zeta}{dt}$.

We will show in appendix \ref{sec:P Comp Law}
 that the connection $\Gamma^{\mu\nu}{}_{\lambda}(p)$ implicitly defined in our framework by (\ref{eq: comp from conn}) can be extracted from
 a ``translated composition law" $\oplus_{\ell}^{[p]}$ in the following way
\begin{equation}\label{Affine from pcomp}
\Gamma^{\mu\nu}{}_{\lambda}(p)=-\frac{\partial}{\partial q_{\mu}}\frac{\partial}{\partial k_{\nu}}\left(q\oplus_{\ell}^{[p]}k\right)_{\lambda}|_{q=k=p}\end{equation}
For this to hold one must define  the translated composition law $\oplus_{\ell}^{[p]}$ through the same kind of construction
that we have described above, but with each momentum associated to the geodesic connecting it to the ``subtraction point'' $p$
instead of the origin.
In formulas this is the prescription that
\begin{equation}\label{eq:translated comp law}
\begin{cases}
\frac{d}{dt}\frac{d}{ds}\gamma^{[p]}_{\lambda}(s,t)+\Gamma^{\mu\nu}{}_{\lambda}(\zeta_{s}(t))\frac{d\zeta_{\mu}^{(s)}(t)}{dt}\frac{d\gamma^{[p]}_{\nu}(s,t)}{ds}=0\\
\gamma^{[p]}(s,0)=\gamma^{(k,p)}(s)\\
\gamma^{[p]}(0,t)=\gamma^{(q,p)}(t)\\
\gamma^{[p]}(1,1)=q\oplus_{\ell}^{[p]}k\\
\frac{d^{2}}{dt^{2}}\zeta_{\lambda}^{(s)}(t)+\Gamma^{\mu\nu}{}_{\lambda}(\zeta^{(s)}(t))\frac{d\zeta_{\mu}^{(s)}(t)}{dt}\frac{d\zeta_{\nu}^{(s)}(t)}{dt}=0\\
\zeta^{(s)}(0)=\gamma^{(q,p)}(s)\\
\zeta^{(s)}(1)=\gamma^{[p]}(s,1)\end{cases}\end{equation}

where $\gamma^{(q,p)}$ is the geodesic defined by:

\begin{equation}
\begin{cases}
\frac{d^{2}}{dt^{2}}\gamma_{\lambda}^{(q,p)}(t)+\Gamma^{\mu\nu}{}_{\lambda}(\gamma^{(q,p)}(t))\frac{d\gamma_{\mu}^{(q,p)}(t)}{dt}\frac{d\gamma_{\nu}^{(q,p)}(t)}{dt}=0\\
\gamma^{(q,p)}(0)=p\\
\gamma^{(q,p)}(1)=q\end{cases}
\label{eq:geodes for pcomp}\end{equation}
and a similar formula holds also for $\gamma^{(k,p)}$.

\subsection{Aside on an ambiguity in the link from composition law to affine
connection}
Since one of the issues that we are keeping in focus concerns the nature of the link between affine connection and
law of composition of momenta it is important for us to stress an aspect of ambiguity of this link
which arises when the composition law is noncommutative,
$p\oplus_{\ell}k\neq k\oplus_{\ell}p$.
This ambiguity is present both with the standard geometric-interpretation prescription, here
coded in Eq.(\ref{eq: Connection Introduction}), and with the novel geometric interpretation we introduced
through Eqs. (\ref{eq: comp from conn})-(\ref{eq: comp is the extr}). The fact here relevant is that the physics of the composition law of course
resides in the composition law itself, not in its derivatives. As a result of this there is no priority
of the definition (\ref{eq: Connection Introduction}), which is
\begin{equation}
\Gamma^{\mu\nu}{}_{\lambda}(p)=-\frac{\partial}{\partial q_{\mu}}\frac{\partial}{\partial k_{\nu}}\left(q\oplus_{\ell}^{[p]}k\right)_{\lambda}|_{q=k=p} \, ,
\label{eq:Connection introd 1}\end{equation}
over the possible alternative 
 \begin{equation}
\Gamma^{\mu\nu}{}_{\lambda}(p)=-\frac{\partial}{\partial k_{\mu}}\frac{\partial}{\partial q_{\nu}}\left(q\oplus_{\ell}^{[p]}k\right)_{\lambda}|_{q=k=p} \, . \label{eq:second conn def}\end{equation}

The same issue can be described by noticing that given the definition (\ref{eq:Connection introd 1}) one still
has the freedom of specifying which one of the upper indices of $\Gamma^{\mu\nu}{}_{\lambda}$
is the differential index. For differential index we mean the index
of the connection that in the expression of the covariant derivative
is paired with the index of the partial derivative. The ambiguity
then is associated with the  possibility of structuring the
covariant derivative either as
$$\nabla_{(1)}^{\lambda}V^{\mu}
=\partial^{\lambda}V^{\mu}-\Gamma^{\lambda\mu}{}_{\nu}V^{\nu}$$
or as
$$\nabla_{(2)}^{\lambda}V^{\mu}=\partial^{\lambda}V^{\mu}
-\Gamma^{\mu\lambda}{}_{\nu}V^{\nu}$$
 where $V^{\mu}$ is an arbitrary vector.

Evidently the same ambiguity also affects the novel prescription which we are here introducing. This can be noticed by looking at
Eq. (\ref{eq: intergeodesics}): just like one can describe $q\oplus_{\ell} k$
in terms of the parallel transport of $\gamma^{(k)}$ along $\gamma^{(q)}$
it would of course be also possible to describe $q\oplus_{\ell} k$ in terms
of the parallel transport of $\gamma^{(q)}$ along $\gamma^{(k)}$.
This means that one can obtain the composition law associated to a given affine connection
by following the same geometrical construction we illustrated in previous
section, but modifying accordingly the boundary conditions in such
a way that the role of $\gamma^{(k)}$ and $\gamma^{(q)}$ is exchanged.
The end result is
\begin{equation}
\begin{cases}
\frac{d}{dt}\frac{d}{ds}\gamma_{\lambda}(s,t)+\Gamma^{\mu\nu}{}_{\lambda}(\zeta_{s}(t))\frac{d\zeta_{\mu}^{(s)}(t)}{dt}\frac{d\gamma_{\nu}(s,t)}{ds}=0\\
\gamma(s,0)=\gamma^{(q)}(s)\\
\gamma(0,t)=\gamma^{(k)}(t)\\
\gamma(1,1)=q\oplus_{\ell} k\\
\frac{d^{2}}{dt^{2}}\zeta_{\lambda}^{(s)}(t)+\Gamma^{\mu\nu}{}_{\lambda}(\zeta^{(s)}(t))\frac{d\zeta_{\mu}^{(s)}(t)}{dt}\frac{d\zeta_{\nu}^{(s)}(t)}{dt}=0\\
\zeta^{(s)}(0)=\gamma^{(q)}(s)\\
\zeta^{(s)}(1)=\gamma(s,1)\end{cases}\label{eq:secondcompfromconn}\end{equation}

Clearly this ambiguity only affects cases where the connection is torsionful. 
Most importantly, it only concerns the geometric interpretation
of the composition law, rather than the composition law itself, and
therefore it has no implications for physics. It is however something that
needs to be dealt with if one is interested in finding general implications
of the geometry of momentum space for physics (since indeed the ambiguity
is such that somewhat different geometries of momentum space could
be associated to the same physical theory). An example of how this
might come to be relevant is provided by the case of studies intending
to establish some general implications of nonmetricity of momentum
space for the physics of the relevant theories: keeping fixed the
on-shellness and the composition law (which specify the physical content
of the theory) one can easily find cases such that the ambiguity here
highlighted involves a possible description in terms of a momentum
space without nonmetricity and a possible description in terms of
a momentum space with nonmetricity.

\section{Differences between the standard and our novel prescription for the geometry of
momentum space\label{sec:Differences}}
Since a large part of what we are here reporting concerns a novel proposal for the geometric interpretation
of momentum-space kinematics, we find appropriate to pause in this section
for some comments on two related issues:\\
$\bullet$ the strength that in general one can expect for the link between the momentum-space
affine connection and the form of the composition law;\\
$\bullet$ the differences between
 the standard geometric-interpretation prescription, here
coded in Eq.(\ref{eq: Connection Introduction}), and the novel geometric interpretation we introduced
through Eqs.  (\ref{eq: comp from conn})-(\ref{eq: comp is the extr}).

We confine our exploration of these issues up to second order in the deformation scale, which already exposes several interesting structures
and allows us to keep the analysis completely explicit. And we assume the composition law has the natural property $k \oplus_{\ell} 0 = k$.
Our starting point then is the following form of the most general composition law at second order in the
deformation scale $\ell$:
\begin{equation}
\left(q\oplus_{\ell}k\right)_{\lambda}=q_{\lambda}+k_{\lambda}+\ell X_{\lambda}^{\alpha\beta}q_{\alpha}k_{\beta}+\frac{\ell^{2}}{2}Y_{\lambda}^{\alpha\beta\gamma}q_{\alpha}q_{\beta}k_{\gamma}
+\frac{\ell^{2}}{2}Z_{\lambda}^{\alpha\beta\gamma}q_{\alpha}k_{\beta}k_{\gamma}\label{eq:General form of comp}\end{equation}
The matrices  $X$, $Y$ and $Z$ are to be determined experimentally.
Evidently we are free to choose any desired geometrical interpretation of composition laws,
since this geometric interpretation is not in itself observable (the physics of the composition law is indeed contained
in these matrices  $X$, $Y$ and $Z$). But still the choice of geometrical interpretation
could have tangible consequences, since the correct choice of geometrical interpretation might prove
to be particularly advantageous for characterizing the observable properties
of the composition law.
A geometrical interpretation of kinematics on momentum space cannot be right or wrong, but could nonetheless be judged
on the basis of its usefulness (or lack thereof).
An important question then is to what extent one can associate an affine connection on momentum
space to one of these
 composition laws.
Is there always a suitable affine connection for any composition law?
And, if so, is there a unique affine connection that ``fits", according to a given prescription of geometrical interpretation?

In exploring these questions one should possibly look for  simple links between properties of the affine connection and properties
 of the composition law. Candidate simple links of this sort could be for example linking  torsion
to noncommutativity of the composition law and/or linking curvature to non-associativity of the composition law. Furthermore
in Refs.\cite{GACarXiv11105081,GACarxiv1111.5643} the possibility has been raised that perhaps the property of
kinematics on momentum space of being (DSR-)relativistic, which is not in general assured \cite{GACarXiv11105081},
might itself admit a geometric description.

Keeping these issues in the background let us start the analysis from contributions to momentum
 composition at first order in the deformation scale.
It is easy to check that, in the notation of  (\ref{eq:General form of comp}), both the previously standard prescription
 and the novel prescription we are here introducing agree on specifying the corresponding leading-order contribution
 to the affine connection as follows
\begin{equation}
\Gamma^{\mu \nu}\,_\lambda(0)=\ell X_\lambda^{\mu \nu}\label{eq:first order con def}\end{equation}
It is very hard to imagine any natural alternative choice for this leading-order relationship.
And notice that the matrix $X$ and the form of the affine connection in the origin
of momentum space, $\Gamma (0)$, have the same number of
components.

The naturalness of the leading-order prescription (\ref{eq:first order con def}) is very clearly not matched
by the situation encountered at second order. In order to see
this let us expand the connection as
\begin{equation}
\Gamma^{\mu\nu}\,_{\lambda}\left(p\right)=\Gamma^{\mu\nu}\,_{\lambda}\left(0\right)
+ p_\theta \partial^{\theta}\Gamma^{\mu\nu}\,_{\lambda}\left(p\right)\big|_{p=0} \label{eq:connection expansion aside}\end{equation}
which also implies that $\partial^{\theta}\Gamma^{\mu\nu}\,_{\lambda}\left(p\right)\big|_{p=0}$ is of second order in the scale $\ell$.

While there appears to be a unique natural link that can be established between the matrix $X$ characteristic of the
leading-order deformation of the composition law and $\Gamma (0)$, one can imagine several ways
for introducing a link between the matrices $Y,Z$ characteristic of the
next-to-leading-order deformation of the composition law and $\partial\Gamma|_{p=0}$.
The first immediate difficulty in this respect
is given by the fact that $Y$ and $Z$ have more degrees of freedom than
$\partial\Gamma$ . In fact, considering for simplicity  the $1+1$ dimensional
case, we note that $Y$ and $Z$ have a total number of 24 independent
components (they are, respectively, symmetric  in the first two
and in the last two indices), while $\partial\Gamma$ can
handle a maximum of 16 components. This means that the space of possible connections is smaller than the space of possible composition laws.
So, one shall inevitably have one of two options:\\
either only a subset of the possible composition laws will admit an associated affine connection\\
or all possible composition laws will admit an associated affine connection but different composition laws
will be mapped into the same connection.

This second option is the one that applies to the ``standard geometric interpretation", here
coded in Eq.(\ref{eq: Connection Introduction}). We can quickly show this, working at second order in $\ell$. We start of course
from (\ref{eq: Connection Introduction}) which we note down again here for convenience:

\begin{equation}
\Gamma^{\mu\nu}{}_{\lambda}(p)=-\frac{\partial}{\partial q_{\mu}}\frac{\partial}{\partial k_{\nu}}\left(q\oplus_{\ell}^{[p]}k\right)_{\lambda}|_{q=k=p}\label{eq:famigerata aside}\end{equation}

Substituting here the general form of the composition law
(\ref{eq:General form of comp}), using the expression (\ref{affinedef}) and with the condition (\ref{eq:first order con def}),
one easily establishes, up to second order in $\ell$, that
\begin{equation}
\Gamma^{\mu\nu}\,_{\lambda}(p) \simeq \Gamma^{\mu\nu}\,_{\lambda}(0)
+\Gamma^{\mu\beta}_{\lambda}(0)\Gamma^{\gamma\nu}\,_{\beta}(0)p_{\gamma}
+\Gamma^{\alpha\nu}\,_{\lambda}(0)\Gamma^{\gamma\mu}\,_{\alpha}(0)p_{\gamma}
-\Gamma^{\alpha\beta}\,_{\lambda}(0)\Gamma^{\mu\nu}\,_{\beta}(0)p_{\alpha}
-\ell^{2}Z_{\lambda}^{\alpha\mu\nu}p_{\alpha} \label{lastfixed}\end{equation}

Then expanding the connection as in (\ref{eq:connection expansion aside})
one gets
\begin{equation}\label{eq:GammaZrelation}
\partial^{\theta}\Gamma^{\mu\nu}\,_{\lambda}\left(p\right)\big|_{p=0}
=\Gamma^{\mu\beta}\,_{\lambda}(0)\Gamma^{\gamma\nu}\,_{\beta}(0)
+\Gamma^{\alpha\nu}\,_{\lambda}(0)\Gamma^{\gamma\mu}\,_{\alpha}(0)
-\Gamma^{\gamma\beta}\,_{\lambda}(0)\Gamma^{\mu\nu}\,_{\beta}(0)-\ell^{2}Z_{\lambda}^{\gamma\mu\nu}\end{equation}

So we see that the possible role played in the composition law by the matrix $Y$ is completely neglected by the
standard prescription (\ref{eq:famigerata aside}). Indeed,
using the above equation, we can rewrite  (\ref{eq:General form of comp}) so that the dependence
of the most general composition law (\ref{eq:General form of comp})
on the connection, according to the standard prescription (\ref{eq:famigerata aside}), is explicit:
\begin{equation}
q\oplus_{\ell}k=q_{\lambda}+k_{\lambda}
-\ell\bar{\Gamma}^{\alpha\beta}\,_{\lambda}q_{\alpha}k_{\beta}+\frac{\ell^{2}}{2}Y_{\lambda}^{\alpha\beta\gamma}q_{\alpha}q_{\beta}k_{\gamma}+\frac{\ell^{2}}{2}\left(\partial^{\theta}\bar{\Gamma}^{\mu\nu}\,_{\lambda}-\bar{\Gamma}^{\beta\theta}\,_{\lambda}\bar{\Gamma}^{\alpha\gamma}\,_{\theta}-\bar{\Gamma}^{\theta\gamma}\,_{\lambda}\bar{\Gamma}^{\alpha\beta}\,_{\theta}
+\bar{\Gamma}^{\alpha\theta}\,_{\lambda}\bar{\Gamma}^{\beta\gamma}\,_{\theta}\right)q_{\alpha}k_{\beta}k_{\gamma}\end{equation}
where we found convenient to render explicit the powers of $\ell$ through the definitions
\begin{equation}
\begin{cases}
\ell\bar{\Gamma}^{\mu\nu}\,_{\lambda}\equiv\Gamma^{\mu\nu}\,_{\lambda}(0) \\
\ell^2\partial^\theta \bar{\Gamma}^{\mu \nu}\,_\lambda\equiv\partial^{\theta}\Gamma^{\mu\nu}\,_{\lambda}\left(p\right)\big|_{p=0}
\end{cases}
\end{equation}

This shows that according to the standard prescription (\ref{eq:famigerata aside}), the composition law is not completely specified by a given connection, as we have the freedom of arbitrarily choosing the 12 components of $Y$.
So the way in which
the difference in the number of components of $(Y,Z)$ and $\Gamma$
is handled within the standard prescription (\ref{eq:famigerata aside})
 is to discard completely the contribution
given by $Y$. And notice that this does not cover even all of the degrees of freedom of $\partial\Gamma$, which are 16:
those are used to specify only the matrix $Z$ with its 12 degrees of freedom. The degrees of freedom of the connection which, according 
to (\ref{eq:famigerata aside}), are not relevant for the description of any composition law
are easily identified , using (\ref{eq:GammaZrelation}) and exploiting the symmetries of $Z$. Indeed
this leads to the requirement
\begin{equation}
\partial^{\theta}\Gamma^{[\mu\nu]}\,_{\lambda}\left(p\right)\big|_{p=0}
=\Gamma^{[\mu\beta}\,_{\lambda}(0)\Gamma^{\gamma\nu]}\,_{\beta}(0)
+\Gamma^{\alpha[\nu}\,_{\lambda}(0)\Gamma^{\gamma\mu]}\,_{\alpha}(0)
-\Gamma^{\gamma\beta}\,_{\lambda}(0)\Gamma^{[\mu\nu]}{}_{\beta}(0)\label{eq:constraintonconnec}\end{equation}
which in 1+1 dimensions is a set of 4 independent equations on $\partial\Gamma$.
So we find that the standard interpretation of the composition law encoded in (\ref{eq:famigerata aside}) 
at second order in the scale $\ell$ provides a map between the set of 
possible $Z$ and the set of connections which satisfy the requirement (\ref{eq:constraintonconnec}). This is a noteworthy 
difference between the standard 
interpretation (\ref{eq:famigerata aside}) and the novel interpretation
here introduced 
in (\ref{eq: comp from conn})-(\ref{eq: comp is the extr}), which does
 not involve any limitation of the form of the connections.

Since part of our focus here is also on the possibility of commutative (and yet deformed) composition laws,
it is important to notice that this peculiar features of the
 standard prescription (\ref{eq:famigerata aside})
 are  also responsible for the fact that symmetric connections
can correspond to noncommutative composition laws.
\begin{equation}
\Gamma^{\mu\nu}\,_{\lambda}(p)=\Gamma^{\nu\mu}\,_{\lambda}(p) \nRightarrow q\oplus_{\ell} k=k\oplus_{\ell} q\end{equation}

Let us now show that the novel prescription of geometrical interpretation of the composition law
which we are here proposing, introduced
through Eqs. (\ref{eq: comp from conn})-(\ref{eq: comp is the extr}), has completely complementary properties with respect to the standard prescription.
Indeed our novel prescription applies only to a subset of composition laws, but for that subset establishes
a one-to-one correspondence between form of the composition law and form of the affine connection.
This is easily seen by making explicit at second order in $\ell$ the map established by our
Eqs.  (\ref{eq: comp from conn})-(\ref{eq: comp is the extr}) obtaining, as we show in the appendix \ref{sec:Second Order Composition Law}:
\begin{equation}
\left(q\oplus_{\ell}k\right)_{\lambda}=q_{\lambda}
+k_{\lambda}-\ell\bar{\Gamma}^{\alpha\beta}\,_{\lambda}q_{\alpha}k_{\beta}
-\frac{\ell^{2}}{2}\partial^{\rho}\bar{\Gamma}^{\alpha\beta}\,_{\lambda}q_{\alpha}k_{\beta}\left(q_{\rho}+k_{\rho}\right)
+\frac{\ell^{2}}{2}\bar{\Gamma}^{\alpha\beta}\,_{\lambda}\bar{\Gamma}^{\gamma\delta}\,_{\alpha}q_{\gamma}k_{\delta}k_{\beta}
+\frac{\ell^{2}}{2}\bar{\Gamma}^{\alpha\beta}\,_{\lambda}\bar{\Gamma}^{\gamma\delta}\,_{\beta}q_{\alpha}q_{\gamma}k_{\delta}
\label{eq:second order cyclic aside}\end{equation}

Since commutative composition laws are one of the main motivations for our analysis,
let us start by noticing that, unlike the standard prescription (\ref{eq:famigerata aside}),
 our new geometric-interpretation prescription, here summarized in (\ref{eq:second order cyclic aside}),
 is such that
symmetric connections are mapped into commutative composition laws:
\begin{equation}
\begin{array}{cc}
\Gamma^{\mu\nu}{}_{\lambda}(p)=\Gamma^{\nu\mu}{}_{\lambda}(p)\\\Downarrow  \\q\oplus_{\ell} k=q_{\lambda}+k_{\lambda}-\ell\bar{\Gamma}^{\alpha\beta}{}_{\lambda}q_{\alpha}k_{\beta}-\frac{\ell^{2}}{2}\partial^{\rho}\bar{\Gamma}^{\alpha\beta}{}_{\lambda}q_{\alpha}k_{\beta}\left(q_{\rho}+k_{\rho}\right)+\frac{\ell^{2}}{2}\bar{\Gamma}^{\alpha\beta}{}_{\lambda}\bar{\Gamma}^{\gamma\delta}{}_{\alpha}q_{\gamma}k_{\delta}k_{\beta}+\frac{\ell^{2}}{2}\bar{\Gamma}^{\alpha\beta}{}_{\lambda}\bar{\Gamma}^{\gamma\delta}{}_{\beta}q_{\alpha}q_{\gamma}k_{\delta}=\\
  =q_{\lambda}+k_{\lambda}-\ell\bar{\Gamma}^{\beta\alpha}{}_{\lambda}q_{\alpha}k_{\beta}-\frac{\ell^{2}}{2}\partial^{\rho}\bar{\Gamma}^{\beta\alpha}{}_{\lambda}q_{\alpha}k_{\beta}\left(q_{\rho}+k_{\rho}\right)+\frac{\ell^{2}}{2}\bar{\Gamma}^{\beta\alpha}{}_{\lambda}\bar{\Gamma}^{\delta\gamma}{}_{\alpha}q_{\gamma}k_{\delta}k_{\beta}+\frac{\ell^{2}}{2}\bar{\Gamma}^{\beta\alpha}{}_{\lambda}\bar{\Gamma}^{\delta\gamma}{}_{\beta}q_{\alpha}q_{\gamma}k_{\delta}=k\oplus_{\ell} q\end{array}\nonumber\end{equation}

From a wider perspective one should notice that, as shown in (\ref{eq:second order cyclic aside}), according to our prescription
the affine connection on momentum space reflects both properties of the matrix $Y$ and properties of the matrix $Z$ (the two matrices
characterizing in our notation the composition law at second order in $\ell$),
whereas as shown above the standard prescription (\ref{eq:famigerata aside})
is such that only the matrix $Z$ leaves a trace in the form of the affine connection.
However, as also noticed above, the combined number of degrees of freedom of the matrices $Y$ and $Z$ is larger
than the number of degrees of freedom of the affine connection at second order in $\ell$. So our prescription
establishes a dependence of the affine connection on both the matrix $Y$ and the matrix $Z$ but cannot code in the affine connection
the most general form of the composition law.
In order to explore this issue we
compare the second-order expansion (\ref{eq:second order cyclic aside}) with the one holding for the most general composition law, Eq.(\ref{eq:General form of comp}), and we obtain from the parts quadratic in $q$ and from the parts quadratic in $k$
the following system of equations
\begin{equation}
\begin{cases}
-\partial^{\gamma}\bar{\Gamma}^{\alpha\beta}{}_{\lambda}q_{\alpha}k_{\beta}q_{\gamma}+\bar{\Gamma}^{\alpha\theta}{}_{\lambda}\bar{\Gamma}^{\gamma\beta}{}_{\theta}q_{\alpha}q_{\gamma}k_{\beta}=Y_{\lambda}^{\alpha\gamma\beta}q_{\alpha}q_{\gamma}k_{\beta}\\
-\partial^{\gamma}\bar{\Gamma}^{\alpha\beta}{}_{\lambda}q_{\alpha}k_{\beta}k_{\gamma}+\bar{\Gamma}^{\theta\beta}{}_{\lambda}\bar{\Gamma}^{\alpha\gamma}{}_{\theta}q_{\alpha}k_{\gamma}k_{\beta}=Z_{\lambda}^{\alpha\beta\gamma}q_{\alpha}k_{\beta}k_{\gamma}\end{cases}\end{equation}
These two equations, taken separately, require
\begin{equation}\label{eq:invertibility system}
\begin{cases}
\partial^{\gamma}\bar{\Gamma}^{\alpha\beta}{}_{\lambda}=-Y_{\lambda}^{\alpha\gamma\beta}+\bar{\Gamma}^{(\alpha\theta}{}_{\lambda}\bar{\Gamma}^{\gamma)\beta}{}_{\theta}+t_{\lambda}^{\alpha\beta\gamma}\\
\partial^{\gamma}\bar{\Gamma}^{\alpha\beta}{}_{\lambda}=-Z_{\lambda}^{\alpha\beta\gamma}+\bar{\Gamma}^{\theta(\beta}{}_{\lambda}\bar{\Gamma}^{\alpha\gamma)}{}_{\theta}+s_{\lambda}^{\alpha\beta\gamma}\end{cases}\end{equation}
where $t$ is antisymmetric in $\alpha$ and $\gamma$ while
$s$ is antisymmetric in $\beta$ and $\gamma$. The fact that not all possible choices of the matrices $Y$ and $Z$ can be
mapped into one of our affine connections is manifest in the fact that these two solutions are not in general
compatible with each other. The solutions will be compatible with each other if and only if one requires that
\begin{equation}
-Y_{\lambda}^{\alpha\gamma\beta}+\bar{\Gamma}^{(\alpha\theta}{}_{\lambda}\bar{\Gamma}^{\gamma)\beta}{}_{\theta}+t^{\alpha\beta\gamma}=-Z_{\lambda}^{\alpha\beta\gamma}+\bar{\Gamma}^{\theta(\beta}{}_{\lambda}\bar{\Gamma}^{\alpha\gamma)}{}_{\theta}+s^{\alpha\beta\gamma}\label{eq:invertibility condition}\end{equation}
We observe that by taking the sum of cyclic permutations of the upper  free indices on both sides
this may be turned into the requirement that
\begin{equation}
Cycl\left(-Y_{\lambda}^{\alpha\gamma\beta}
+\bar{\Gamma}^{(\alpha\theta}{}_{\lambda}\bar{\Gamma}^{\gamma)\beta}{}_{\theta}\right)
=Cycl\left(-Z_{\lambda}^{\alpha\beta\gamma}
+\bar{\Gamma}^{\theta(\beta}{}_{\lambda}\bar{\Gamma}^{\alpha\gamma)}{}_{\theta}\right)
\label{eq:cycliccondition}\end{equation}
where we introduced $Cycl$ defined as
\begin{equation}
Cycl\left(W_{\lambda}^{\alpha\beta\gamma}\right) \equiv W_{\lambda}^{\alpha\beta\gamma}+W_{\lambda}^{\gamma\alpha\beta}
+W_{\lambda}^{\beta\gamma\alpha}\end{equation}

Furthermore, it takes some simple algebra to verify that, 
if the cyclic condition (\ref{eq:cycliccondition}) holds, then a solution of (\ref{eq:invertibility condition}) exists 
and is given by: 
\begin{equation}
\begin{cases}
t^{\alpha\beta\gamma}=\frac{2}{3}\tilde{Y}^{[\alpha\beta\gamma]}-\frac{4}{3}\tilde{Z}^{[\alpha\beta\gamma]}\\
s^{\alpha\beta\gamma}=\frac{2}{3}\tilde{Z}^{[\gamma\alpha\beta]}-\frac{4}{3}\tilde{Y}^{[\gamma\alpha\beta]}\end{cases}\label{eq:tesform}\end{equation}
where we defined
\begin{equation}
\begin{cases}
\tilde{Y}_{\lambda}^{\alpha\beta\gamma}=Y_{\lambda}^{\alpha\beta\gamma}-\Gamma^{(\alpha\theta}{}_{\lambda}\Gamma^{\beta)\gamma}{}_{\theta}\\
\tilde{Z}_{\lambda}^{\alpha\beta\gamma}=Z_{\lambda}^{\alpha\beta\gamma}-\Gamma^{\theta(\gamma}{}_{\lambda}\Gamma^{\alpha\beta)}{}_{\theta}\end{cases}\end{equation}
  Inserting the expressions (\ref{eq:tesform}) for $s$ and $t$  into the system of equations (\ref{eq:invertibility system}) and recalling (\ref{eq:connection expansion aside}) we find that the connection as a function of $Y$ and $Z$ can be given in the forms :
\begin{equation}\label{eq:ourinversemap}
\begin{cases}
\Gamma^{\alpha\beta}{}_{\lambda}\left(p\right)=\ell\bar{\Gamma}^{\alpha\beta}{}_{\lambda}+\ell^2\left(-\tilde{Y}_{\lambda}^{\alpha\gamma\beta}+\frac{2}{3}\tilde{Y}^{[\alpha\beta\gamma]}-\frac{4}{3}\tilde{Z}^{[\alpha\beta\gamma]}\right)p_{\gamma}\\
\Gamma^{\alpha\beta}{}_{\lambda}\left(p\right)=\ell\bar{\Gamma}^{\alpha\beta}{}_{\lambda}+\ell^2\left(-\tilde{Z}_{\lambda}^{\alpha\beta\gamma}+\frac{2}{3}\tilde{Z}^{[\gamma\alpha\beta]}-\frac{4}{3}\tilde{Y}^{[\gamma\alpha\beta]}\right)p_{\gamma}\end{cases}
\end{equation}
which are evidently equivalent to each other because of (\ref{eq:cycliccondition}).
These (\ref{eq:ourinversemap})
are the analogue, for our construction of the composition law, of equation (\ref{eq: Connection Introduction}); 
they allow us to associate at second order in $\ell$ a connection $\Gamma$ to any given composition law satisfying the 
cyclic condition (\ref{eq:cycliccondition}).

We can characterize the cyclic condition (\ref{eq:cycliccondition}) more vividly by rewriting
it as the following property that,
at quadratic order\footnote{Eq.(\ref{eq:cyclic identity}) is easily verified by expanding it at second order
in $\ell$. Also note that since Eq.(\ref{eq:cycliccondition}) characterizes only a second-order-in-$\ell$ condition,
also Eq.(\ref{eq:cyclic identity}) must be viewed as here determined only up to quadratic order in $\ell$ (in particular one could 
replace $\ominus_{\ell} q$
with $\oplus_{\ell} q$ in (\ref{eq:cyclic identity}) 
and still get a relation  equivalent to (\ref{eq:cycliccondition}) at quadratic order in $\ell$).} in $\ell$, needs
to be imposed on the composition law:
\begin{equation}
Cycl_{\oplus_{\ell}}\left\{ \left(\ominus_{\ell} q\oplus_{\ell}\left(k\oplus_{\ell} q\right)\right)_{\lambda}\right\} =Cycl_{\oplus_{\ell}}\left\{ \left(\left(\ominus_{\ell} q\oplus_{\ell} k\right)\oplus_{\ell} q\right)_{\lambda}\right\} \label{eq:cyclic identity}\end{equation}
where $Cycl_{\oplus_{\ell}}$ is  defined by
\begin{equation}
\begin{cases}
Cycl_{\oplus_{\ell}} \left\{ \left(p\oplus_{\ell}\left(k\oplus_{\ell} q\right)\right)_{\lambda}\right\} =\left(p\oplus_{\ell}\left(k\oplus_{\ell} q\right)\right)_{\lambda}+\left(q\oplus_{\ell}\left(p\oplus_{\ell} k\right)\right)_{\lambda}+\left(k\oplus_{\ell}\left(q\oplus_{\ell} p\right)\right)_{\lambda}\\
Cycl_{\oplus_{\ell}}\left\{ \left(\left(p\oplus_{\ell} k\right)\oplus_{\ell} q\right)_{\lambda}\right\} =\left(\left(p\oplus_{\ell} k\right)\oplus_{\ell} q\right)_{\lambda}+\left(\left(q\oplus_{\ell} p\right)\oplus_{\ell} k\right)_{\lambda}+\left(\left(k\oplus_{\ell} q\right)\oplus_{\ell} p\right)_{\lambda}\end{cases}\end{equation}

We also observe that, focusing again on the 1+1D case, one has that
the condition for sums of cyclic permutations (\ref{eq:cycliccondition}) corresponds to a
set of 8 independent conditions. The matrices $Y$ and $Z$ start off with a total of 24 degrees of freedom, and
after this 8 conditions we are left with 16 degrees of freedom which exactly matches the number of degrees
of freedom of the quadratic-in-$\ell$ part of the affine connection. This should be contrasted to the fact observed above that
only 12 degrees of freedom of the composition law (the ones of the matrix $Z$) leave a trace in the affine connection
according to the standard prescription  (\ref{eq:famigerata aside}).

\section{Two $\kappa$-de Sitter momentum spaces}

The possibility which has been so far most
studied \cite{GiuliaFlavio,Trevisan,MercatiCarmonaCortes,LinQing}
 within the relative-locality framework is the one
of the $\kappa$-dS ({}``$\kappa$-de Sitter\textquotedbl{}) momentum
space. This is based on a form of on-shellness and a form of the law
of composition of momenta inspired by the $k$-Poincar\'e Hopf algebra
\cite{majrue,lukieANNALS}, which had already been of interest from
the quantum-gravity perspective for independent reasons \cite{gacPLB1997,gacmajid,freidellivinePRL,leeCURVEDMOMENTUM}.

As most previous works on this  possibility \cite{GiuliaFlavio,Trevisan}
we shall focus on the 1+1-dimensional $\kappa$-dS momentum space,
which allows us to discuss the key conceptual features in a slightly
simplified context (with respect to the 3+1-dimensional case). For the 1+1-dimensional $\kappa$-dS momentum space the metric is
a de Sitter metric, \begin{equation}
g^{\mu\nu}(p)=\left(\begin{array}{cc}
1 & 0\\
0 & -e^{2\ell p_{0}}\end{array}\right)\end{equation}
 while the composition law has the form \cite{GiuliaFlavio,Trevisan}
\begin{equation}
\begin{cases}
\left(q\oplus_{\ell}k\right)_{0}=q_{0}+k_{0}\\
\left(q\oplus_{\ell}k\right)_{1}=q_{1}+k_{1}e^{-\ell p_{0}}\end{cases}\label{eq:k composition}\end{equation}
We note down the antipode of this composition law
\[
\begin{cases}
\left(p\oplus_{\ell}\left(\ominus_{\ell}p\right)\right)_{0}=p_{0}+\ominus_{\ell}p_{0}=0\rightarrow\ominus_{\ell}p_{0}=-p_{0}\\
\left(p\oplus_{\ell}\left(\ominus_{\ell}p\right)\right)_{1}=p_{1}+\ominus_{\ell}p_{1}e^{-\ell p_{0}}=0\rightarrow\ominus_{\ell}p_{1}=-p_{1}e^{\ell p_{0}}\end{cases}\]
which in fact is such that $\ominus_{\ell} p \oplus_{\ell} p =0 = p\oplus_{\ell} (\ominus_{\ell} p)$.

And we notice that the $\kappa$-dS composition law is associative
\begin{equation}
(p\oplus_{\ell} q)\oplus_{\ell} k=p\oplus_{\ell} (q\oplus_{\ell} k) \, .
\end{equation}
For reasons that will be clearer in the following let us also observe that evidently for associative composition
laws one has that the so-called ``left loop inverse rule" and ``right loop inverse rule" \cite{Freidel:2011mt} apply:
\begin{equation}
q\oplus_{\ell}\left(\ominus_{\ell} q\oplus_{\ell} p\right)=p\label{eq:leftloop}\end{equation}
\begin{equation}
\left(p\oplus_{\ell}\ominus_{\ell} q\right)\oplus_{\ell} q=p\label{eq:rightloop}.\end{equation}

Using the prescription summarized in the previous section (Eq. (\ref{eq: On-shell is the length}))
one easily finds that the de Sitter metric on momentum space leads
to the following on-shell relation:
 \begin{equation}
m^2=d_{\ell}\left(p,0\right)=\frac{1}{\ell}\text{Arccosh}\left[\text{Cosh}\left[\ell p_{0}\right]-\frac{\ell^{2}}{2}p_{1}^{2}e^{\ell p_{0}}\right]\end{equation}
We notice that this too matches results in the $\kappa$-Poincar\'e literature,
since it is evidently related to the "$\kappa$-deformed mass Casimir" customarily written as
 \begin{equation}
\text{Cosh}[\ell m]=\text{Cosh}[\ell d_{\ell}\left(p\right)]=\text{Cosh}[\ell p_{0}]-\frac{\ell^{2}}{2}p_{1}^{2}e^{\ell p_{0}}\end{equation}

This $\kappa$-dS momentum space is here of interest from several viewpoints. It is the most studied curved momentum space, and yet
we have several points to contribute to its understanding (particularly concerning its relativistic properties, see later).
Its composition law is noncommutative (as for all other momentum spaces so far studied in some detail), and so it serves
as a reference of contrast to our proposal of a de Sitter momentum space with commutative composition law.
And importantly this is an example of momentum space whose description is within reach of both
 the standard geometric-interpretation prescription, here
coded in Eq.(\ref{eq: Connection Introduction}), and the novel geometric interpretation we introduced
through Eqs.  (\ref{eq: comp from conn})-(\ref{eq: comp is the extr}).
Since in this manuscript we often stress the differences between the standard geometric-interpretation prescription and our
 novel geometric interpretation it is important for us to also provide an example where no such differences are present.
If all we were interested in were only cases like the $\kappa$-dS momentum space then there would be no
difference between the standard geometric interpretation  and our
 novel geometric interpretation.

For what concerns the description of the ``$\kappa$-connection", {\it i.e.} the affine connection for the
$\kappa$-dS momentum space, as obtained adopting the standard geometric-interpretation prescription, here
coded in Eq.(\ref{eq: Connection Introduction}), we can rely on the previous analyses of Refs.~\cite{GiuliaFlavio,Trevisan}
establishing that
 \begin{equation}
\Gamma^{\lambda\mu}{}_{\nu}
=\ell\delta_{0}^{\lambda}\delta_{1}^{\mu}\delta_{\nu}^{1}
\label{eq:k-connection}\end{equation}
This result can be quickly reproduced starting from Eq.(\ref{eq: Connection Introduction}), which is
\begin{equation}\Gamma^{\mu\nu}{}_{\lambda}(p)=-\frac{\partial}{\partial q_{\mu}}\frac{\partial}{\partial k_{\nu}}\left(q\oplus_{\ell}^{[p]}k\right)_{\lambda}\Big|_{q=k=p}\end{equation}
where in the $\kappa$-dS case we have that
\begin{equation}
\begin{cases}
\left(q\oplus_{\ell}^{[p]}k\right)_{0}=\left(p\oplus_{\ell}\left[\left(\ominus_{\ell}\, p\oplus_{\ell}q\right)\oplus_{\ell}\left(\ominus_{\ell}\, p\oplus_{\ell}k\right)\right]\right)_0=q_{0}+k_{0}-p_{0}\\
\left(q\oplus_{\ell}^{[p]}k\right)_{1}=\left(p\oplus_{\ell}\left[\left(\ominus_{\ell}\, p\oplus_{\ell}q\right)\oplus_{\ell}\left(\ominus_{\ell}\, p\oplus_{\ell}k\right)\right]\right)_1=p_{1}+\left(q_{1}+k_{1}-2p_{1}\right)e^{-\ell\left(q_{0}-p_{0}\right)}\end{cases}\end{equation}
From this it follows that
\begin{equation}
\Gamma^{\mu\nu}{}_{\lambda}(p)=-\frac{\partial}{\partial q_{\mu}}\frac{\partial}{\partial k_{\nu}}\left(q\oplus_{\ell}^{[p]}k\right)_{\lambda}\Big|_{q=k=p}=\ell\delta_{0}^{\mu}\delta_{1}^{\nu}\delta_{\lambda}^{1}\end{equation}

 It is interesting to notice that this 1+1-dimensional $\kappa$-dS momentum space has de Sitter metric (constant
curvature $- 2 \ell^{2}$) and affine connection which instead has no curvature:
 \begin{equation}
F^{\mu\beta\alpha}{}_{\lambda}=\partial^{\mu}\Gamma^{\beta\alpha}\,_{\lambda}
-\partial^{\beta}\Gamma^{\mu\alpha}\,_{\lambda}
+\Gamma^{\mu\tau}\,_{\lambda}\Gamma^{\beta\alpha}\,_{\tau}
-\Gamma^{\beta\tau}\,_{\lambda}\Gamma^{\mu\alpha}\,_{\tau}=0\end{equation}

\subsection{$\kappa$-dS composition law from $\kappa$-connection}

Our first task is to verify that the $\kappa$-connection
(\ref{eq:k-connection}) does produce the $\kappa$-dS composition
law when applying the geometric construction of the composition law
encoded in Eqs.  (\ref{eq: comp from conn})-(\ref{eq: comp is the extr}).

We start by noticing that the equation (\ref{eq: comp from conn}),
which defines the composition law associated to a given affine connection,
in the case of the $\kappa$-connection takes the form: \begin{equation}
\begin{cases}
\frac{d}{dt}\frac{d}{ds}\gamma_{0}(s,t)=0\\
\frac{d}{dt}\frac{d}{ds}\gamma_{1}(s,t)+\ell\frac{d\zeta_{0}(s,t)}{dt}\frac{d\gamma_{1}(s,t)}{ds}=0\end{cases}\end{equation}
 where $\zeta$ is the connection geodesic connecting $\gamma(s,0)$
to $\gamma(s,1)$ , which means \begin{equation}
\begin{cases}
\frac{d^{2}}{dt^{2}}\zeta_{0}(s,t)=0\\
\frac{d^{2}}{dt^{2}}\zeta_{1}(s,t)+\ell\frac{d\zeta_{0}(s,t)}{dt}\frac{d\zeta_{1}(s,t)}{dt}=0\\
\zeta_{1}(s,0)=\gamma(s,0)\\
\zeta_{1}(s,1)=\gamma(s,1)\end{cases}\end{equation}

The solution for $\zeta$ is easily found to take the form \begin{equation}
\begin{cases}
\zeta_{0}(s,t)=\gamma_{0}(s,0)+\Delta_{0}t\\
\zeta_{1}(s,t)=\gamma_{1}(s,0)+\Delta_{1}\frac{\left(1-e^{-\ell\Delta_{0}t}\right)}{\left(1-e^{-\ell\Delta_{0}}\right)}\end{cases}\end{equation}
 where \begin{equation}
\Delta_{\lambda}=\gamma_{\lambda}(s,1)-\gamma_{\lambda}(s,0)\end{equation}

Using this solution one then finds that $\gamma$ must satisfy \begin{equation}
\begin{cases}
\frac{d}{dt}\frac{d}{ds}\gamma_{0}(s,t)=0\\
\frac{d}{dt}\frac{d}{ds}\gamma_{1}(s,t)+\ell\Delta_{0}\frac{d\gamma_{1}(s,t)}{ds}=0\end{cases}\end{equation}

In turn these, once the boundary conditions (\ref{eq: comp boundary conditions})
are imposed on $\gamma(s,t)$, lead to the results \begin{equation}
\begin{cases}
\gamma_{0}(s,t)=\gamma_{0}^{(k)}(s)+\gamma_{0}^{(q)}(t)\\
\gamma_{1}(s,t)=\gamma_{1}^{(q)}(t)+\gamma_{1}^{(k)}(s)e^{-\ell q_{0}t}\end{cases}\end{equation}
 where $\gamma^{(q)}$ ( $\gamma^{(k)}$ ) is the connection geodesic
connecting the point $q$ (the point $k$) to the origin.

Then, using our formulation of the composition law given in terms
of $q\oplus_{\ell} k=\gamma(1,1)$, one easily finds that
\begin{equation}\label{eq:k comp law}
\begin{cases}
\left(q\oplus_{\ell} k\right)_{0}=q_{0}+k_{0}\\
\left(q\oplus_{\ell} k\right)_{1}=q_{1}+k_{1}e^{-\ell q_{0}}\end{cases}\end{equation}
 which indeed successfully reproduces the $\kappa$-dS composition
law.

Following steps completely analogous to the ones discussed in this
subsection it is easy to verify that one obtains again correctly the
$\kappa$-dS composition law also using the alternative definition
for the affine connection given in (\ref{eq:second conn def}) (and
of course then using the corresponding steps of derivation, summarized
in the previous section in Eqs. (\ref{eq:secondcompfromconn})).

\subsection{Two $\kappa$-dS geometries}

The last point we want to make in this section on the $\kappa$-dS
momentum space concerns again the ambiguity we highlighted for what
concerns the association of an affine connection to a given law of
composition of momenta. We recall here that the ambiguity is connected
with the choice between the following two options: \begin{equation}
\begin{cases}
\Gamma_{(1)}^{\lambda\mu}{}_{\nu}(p)=-\frac{\partial}{\partial q_{\lambda}}\frac{\partial}{\partial k_{\mu}}\left(q\oplus_{\ell}^{[p]}k\right)_{\nu}|_{q=k=p}\\
\Gamma_{(2)}^{\lambda\mu}{}_{\nu}(p)=-\frac{\partial}{\partial k_{\lambda}}\frac{\partial}{\partial q_{\mu}}\left(q\oplus_{\ell}^{[p]}k\right)_{\nu}|_{q=k=p}\end{cases}\label{eq: splitting of geometries}\end{equation}

As stressed at the end of the previous subsection, in the $\kappa$-dS
case one can explicitly verify that both of these possibilities for
the affine connection give us back the $\kappa$-dS composition law.
This confirms that the choice between $\Gamma_{(1)}$ and $\Gamma_{(2)}$
is merely conventional. But it is still interesting to assess how
this ambiguity affects the geometric structure of $\kappa$-dS momentum
space. As a step toward doing this let us start by recalling that
(as reviewed in the appendix \ref{sec:Decomposition of the Connection}),
given a metric, any affine connection can be split in a unique way
as follows:
\begin{equation}
\Gamma^{\lambda\mu}{}_{\nu}=A^{\lambda\mu}{}_{\nu}+K^{\lambda\mu}{}_{\nu}+V^{\lambda\mu}{}_{\nu}
\end{equation}
 where $K$ is the contortion tensor, defined by
\begin{equation}
K^{\lambda\mu}{}_{\rho}g^{\rho\nu}+K^{\lambda\nu}{}_{\rho}g^{\rho\mu}
=0\end{equation}
 and $V$ is the cononmetricity, defined by \begin{equation}
\nabla_{(\Gamma)}^{\lambda}g^{\mu\nu}=
V^{\lambda\mu}{}_{\rho}g^{\rho\nu}+V^{\lambda\nu}{}_{\rho}g^{\rho\mu}\end{equation}
Next let us observe that for the $\kappa$-dS composition law (\ref{eq:k composition})
one finds that
\begin{equation}
\begin{cases}
\Gamma_{(1)}^{\lambda\mu}{}_{\nu}
=\ell\delta_{0}^{\lambda}\delta_{1}^{\mu}\delta_{\nu}^{1}\\
\Gamma_{(2)}^{\lambda\mu}{}_{\nu}=\ell\delta_{0}^{\mu}\delta_{1}^{\lambda}\delta_{\nu}^{1}\end{cases}\end{equation}
 while the Levi-Civita connection on de Sitter momentum space is \begin{equation}
A^{\lambda\mu}{}_{\nu}=\ell\left(\delta_{0}^{\lambda}\delta_{1}^{\mu}+\delta_{1}^{\lambda}\delta_{0}^{\mu}\right)\delta_{\nu}^{1}+\ell e^{2\ell p_{0}}\delta_{1}^{\lambda}\delta_{1}^{\mu}\delta_{\nu}^{0}\end{equation}

By subtracting this $A^{\lambda\mu}{}_{\nu}$ to both $\Gamma_{(1)}$
and $\Gamma_{(2)}$ we can get the contortion and the cononmetricity
in the two cases:\\
\begin{equation}
\begin{cases}
K_{(1)}^{\mu\nu}\,_{\lambda}=-\ell \delta^{\mu}_{1}\left( \delta^{\nu}_{1}\delta^{0}_{\lambda}e^{2\ell p_{0}}+\delta_{\nu}^{0}\delta^{1}_{\lambda}\right) &\quad \quad K_{(2)}^{\mu\nu}\,_{\lambda}=-K_{(1)}^{\mu\nu}\,_{\lambda}\\
V_{(1)}^{\mu\nu}\,_{\lambda}=0 &\quad \quad V_{(2)}^{\mu\nu}\,_{\lambda}=2K_{(1)}^{\mu\nu}\,_{\lambda}\end{cases}\end{equation}

Notice in particular that $\Gamma_{(1)}$ has no cononmetricity whereas
for $\Gamma_{(2)}$ the cononmetricity does not vanish. This illustrates
the issue we raised above: results on the relative-locality framework
with the $\kappa$-dS momentum space cannot be used for developing
any general intuition on the role of cononmetricity in this sort of
theories, since the $\kappa$-dS theory can be viewed in equally legitimate
manner both as a case of momentum space with cononmetricity and as
a case of momentum space without cononmetricity.

\section{{}``Proper\textquotedbl{} de Sitter momentum space}

The $\kappa$-dS momentum space has been a preferred choice for the
first studies done within the relative-locality framework also for
the simplifications afforded by the $\kappa$-connection and corresponding
simple properties of the composition law, which in particular is associative.
The ``price to pay" for that associativity is the noncommutativity of the composition
law on the $\kappa$-momentum space.
One of the main objectives of the study we are here reporting is to
propose a momentum space which could be the most natural starting
point for the first investigations of the relative-locality framework
when the composition law is not associative but it is commutative. The case we propose for
these purposes is {}``proper\textquotedbl{} de Sitter momentum space
(which we label alternatively as the proper-dS momentum space or simply as the dS momentum space). By this
we mean a momentum space whose metric is a de Sitter metric, like
in the $\kappa$-dS case, and as the affine connection on momentum
space one takes the Levi-Civita connection of the de Sitter metric
(rather than the $\kappa$-connection). So we deal with the same metric
as in the previous section, \begin{equation}
g^{\mu\nu}(p)=\left(\begin{array}{cc}
1 & 0\\
0 & -e^{2\ell p_{0}}\end{array}\right)\end{equation}
 but now studied in combination with its associated Levi-Civita connection
 \begin{equation}
A^{\lambda\mu}{}_{\nu}
=\ell\left(\delta_{0}^{\lambda}\delta_{1}^{\mu}
+\delta_{0}^{\mu}\delta_{1}^{\lambda}\right)\delta_{\nu}^{1}
+\ell e^{2\ell p_{0}}\delta_{1}^{\lambda}\delta_{1}^{\mu}\delta_{\nu}^{0}\label{eq:Levi Civita de sitter connection}\end{equation}
 rather than the $\kappa$-connection.

Since the on-shell relation is determined exclusively by the metric,
for theories on this proper-dS momentum space one must enforce the same on-shellness
as in the $\kappa$-dS case: \begin{equation}
\text{Cosh}[\ell m]=\text{Cosh}[\ell d_{\ell}\left(p\right)]=\text{Cosh}[\ell p_{0}]-\frac{\ell^{2}}{2}p_{1}^{2}e^{\ell p_{0}}\end{equation}

We find that this setup leads to a commutative (but non-associative) composition law when
assuming for the link between composition law and affine connection the novel geometric-interpretation
prescription we introduced in Eqs.(\ref{eq: comp from conn})-(\ref{eq: comp is the extr}).

Contrary to the $\kappa$-momentum case, in this case of combining the de Sitter metric with its Levi-Civita connection
one does find some  differences between
 the standard geometric-interpretation prescription, here
coded in Eq.(\ref{eq: Connection Introduction}), and the novel geometric interpretation we introduced
through Eqs.  (\ref{eq: comp from conn})-(\ref{eq: comp is the extr}).
When the analysis is done with the prescription (\ref{eq: Connection Introduction}) the composition law associated
to the Levi-Civita connection is not necessarily commutative, but, in a sense that we shall soon clarify,
 one can choose to make it commutative. Even when that particular choice of commutativity of the composition law
 is made the resulting composition law is different from the one obtained from the novel geometric-interpretation
prescription we introduced in Eqs.(\ref{eq: comp from conn})-(\ref{eq: comp is the extr}).

\subsection{Weakly-proper dS\label{sec:Improper dS}}
One of our main goals is to find a natural candidate
for a commutative composition law on a momentum space with dS metric, and our
preferred scenario when this project got started was to find that such a picture
could be directly linked to the adoption of the Levi-Civita connection on dS momentum space.
In the next subsection we shall report a picture which exactly matches these desiderata and expectations,
relying on the
novel geometric-interpretation
prescription we introduced in Eqs.(\ref{eq: comp from conn})-(\ref{eq: comp is the extr}).
We shall label that picture as ``proper dS momentum space".
Before getting to those results, in this subsection we show that also adopting the standard geometric-interpretation prescription, here
coded in Eq.(\ref{eq: Connection Introduction}), one can have on dS momentum space a commutative composition law paired
with the Levi-Civita connection, but the logical link between these two structures is weaker, in the sense already here
clarified in Sec.III (and visible again in the points we make in this subsection).
We shall then label the picture arising in this subsection as the ``weakly-proper dS momentum space".

Our road to this weakly-proper dS momentum space starts of course again from (\ref{eq: Connection Introduction}):
\begin{equation}\Gamma^{\mu\nu}{}_\lambda(p)=-\frac{\partial}{\partial q_{\mu}}\frac{\partial}{\partial k_{\nu}}\left(q\oplus_{\ell}^{[p]}k\right)_{\lambda}|_{q=k=p}\end{equation}

Using notation and results from Sec.III, now specialized to the case of the
Levi-Civita connection on dS momentum space, (\ref{eq:Levi Civita de sitter connection}),
one easily finds that
\begin{equation}
X_{\lambda}^{\mu\nu}=\left(\delta_{0}^{\lambda}\delta_{1}^{\mu}+\delta_{0}^{\mu}\delta_{1}^{\lambda}\right)\delta_{\nu}^{1}+\delta_{1}^{\lambda}\delta_{1}^{\mu}\delta_{\nu}^{0}\end{equation}
and
\begin{equation}
Z_{\lambda}^{\alpha\mu\nu}=\ell^{2}\left[\delta_{0}^{\mu}\delta_{\lambda}^{1}\delta_{0}^{\nu}\delta_{1}^{\alpha}
+\frac{1}{2}\delta_{1}^{\mu}\delta_{\lambda}^{1}\delta_{1}^{\alpha}\delta_{1}^{\nu}\right]\end{equation}
So the family of composition laws that can be
 associated to the de Sitter Levi-Civita connection on dS momentum space
according to the standard geometric-interpretation prescription, here
coded in Eq.(\ref{eq: Connection Introduction}), takes the form:
\begin{equation}
\left(q\oplus_{\ell} k\right)_{\lambda}=q_{\lambda}+k_{\lambda}
-A^{\alpha\beta}{}_{\lambda}\left(0\right)q_{\alpha}k_{\beta}
+\ell^{2}Y_{\lambda}^{\alpha\beta\gamma}q_{\alpha}q_{\beta}k_{\gamma}
+\ell^{2}\delta_{\lambda}^{1}\left(q_{1}k_{0}^{2}
+\frac{1}{2}q_{1}k_{1}^{2}\right)q_{\alpha}k_{\mu}k_{\nu}
\label{jocjoc}
\end{equation}
where $A_{\lambda}^{\alpha\beta}$ is the de Sitter Levi-Civita connection, whose form was noted in (\ref{eq:Levi Civita de sitter connection}),
 and $Y_{\lambda}^{\alpha\beta\gamma}$ is an arbitrary constant
tensor. Eq.(\ref{jocjoc}) can be rewritten more explicitly as follows 
\begin{equation}\label{eq:improper class}
\begin{cases}
\left(q\oplus_{\ell}k\right)_{0}=q_{0}+k_{0}-\ell q_{1}k_{1}
+\ell^{2}Y_{0}^{\alpha\beta\gamma}q_{\alpha}q_{\beta}k_{\gamma}\\
\left(q\oplus_{\ell}k\right)_{1}=q_{1}+k_{1}-\ell\left(q_{0}k_{1}
+q_{1}k_{0}\right)+\ell^{2}Y_{1}^{\alpha\beta\gamma}q_{\alpha}q_{\beta}k_{\gamma}
+\ell^{2}\left(q_{1}k_{0}^{2}+\frac{1}{2}q_{1}k_{1}^{2}\right)\end{cases}\end{equation}

Evidently this family of composition laws includes a possibility which is commutative,
and is therefore of particular interest for some of the objectives of our analysis. This commutative
composition law is
\begin{equation}
\begin{cases}
\left(q\oplus_{\ell}k\right)_{0}=q_{0}+k_{0}-\ell q_{1}k_{1}\\
\left(q\oplus_{\ell}k\right)_{1}=q_{1}+k_{1}-\ell\left(q_{0}k_{1}+q_{1}k_{0}\right)
+\ell^{2}\left(q_{1}k_{0}^{2}+q_{0}^{2}k_{1}
+\frac{1}{2}q_{1}k_{1}^{2}+\frac{1}{2}q_{1}^2k_{1}\right)\end{cases}\label{jocccc}\end{equation}

We shall label as weakly-proper dS momentum space the case characterized (to quadratic order in $\ell$)
 by this commutative composition law
and the on-shellness determined by the dS metric
on momentum space.

We observe that the composition law (\ref{jocccc}) is not associative, 
\begin{equation}
\left(\left(q\oplus_{\ell}k\right)\oplus_{\ell}p\right)_{\lambda}
\neq\left(q\oplus_{\ell}\left(k\oplus_{\ell}p\right)\right)_{\lambda}
\end{equation}
and one easily verifies that it does not satisfy neither the left loop rule nor the right loop rule:
\begin{equation}
q\oplus_{\ell}\left(\ominus_{\ell} q\oplus_{\ell} p\right)\neq p\end{equation}
\begin{equation}
\left(p\oplus_{\ell}\ominus_{\ell} q\right)\oplus_{\ell} q \neq p.\end{equation}

\subsection{Proper dS}
Our next task is to show that
the novel geometric-interpretation
prescription we introduced in Eqs.(\ref{eq: comp from conn})-(\ref{eq: comp is the extr})
associates the Levi-Civita connection on momentum space to a specific commutative composition law.

We show in appendix \ref{sec:Second Order Composition Law} how to
derive from our Eqs.(\ref{eq: comp from conn})-(\ref{eq: comp is the extr})
a second-order-in-$\ell$ expression for the composition law. This requires  expanding the connection as follows:
\begin{equation}
\Gamma^{\lambda\mu}{}_{\nu}(\zeta)=\Gamma^{\lambda\mu}{}_{\nu}(0)
+\partial^{\rho}\Gamma^{\lambda\mu}{}_{\nu}(0)\zeta_{\rho}^{(0)}\end{equation}
 where we defined $\zeta^{(0)}$ as the zero-th order approximation
of $\zeta$ (definitions in Eqs.(\ref{eq: comp from conn})-(\ref{eq: comp is the extr})).

Adopting analogous expansions for $\gamma$ and $\zeta$ one then obtains with
simple steps of derivation the result
\begin{equation}
\left(q\oplus_{\ell}k\right)_{\lambda}=q_{\lambda}+k_{\lambda}
-\ell\bar{\Gamma}^{\alpha\beta}{}_{\lambda}q_{\alpha}k_{\beta}
-\ell^{2}\left(\partial^{\rho}\bar{\Gamma}^{\alpha\beta}{}_{\lambda}\right)q_{\alpha}k_{\beta}\left(\frac{q_{\rho}+k_{\rho}}{2}\right)
+\frac{1}{2}\ell^{2}\bar{\Gamma}^{\alpha\beta}{}_{\lambda}\bar{\Gamma}^{\gamma\delta}{}_{\alpha}q_{\gamma}k_{\delta}k_{\beta}
+\frac{1}{2}\ell^{2}\bar{\Gamma}^{\alpha\beta}{}_{\lambda}\bar{\Gamma}^{\gamma\delta}{}_{\beta}q_{\alpha}q_{\gamma}k_{\delta}
\label{eq:2 order comp law}\end{equation}
 where again we use the notation  $\ell\bar{\Gamma}^{\mu\nu}{}_{\lambda}\equiv \Gamma^{\mu\nu}{}_{\lambda}(0)$ , $\ell^2\partial^\theta \bar{\Gamma}^{\mu \nu}{}_\lambda\equiv\partial^{\theta}\Gamma^{\mu\nu}{}_{\lambda}\left(p\right)\big|_{p=0}$.
Substituting in this result the explicit form of the Levi-Civita connection
(\ref{eq:Levi Civita de sitter connection}) one then finally obtains
\begin{equation}
\begin{cases}
\left(q\oplus_{\ell}k\right)_{0}=q_{0}+k_{0}-\ell q_{1}k_{1}+\frac{\ell^{2}}{2}\left[-q_{1}k_{1}\left(q_{0}+k_{0}\right)+q_{0}k_{1}^{2}+q_{1}^{2}k_{0}\right]\\
\left(q\oplus_{\ell}k\right)_{1}=q_{1}+k_{1}-\ell\left(q_{0}k_{1}+q_{1}k_{0}\right)+\frac{\ell^{2}}{2}\left[\left(q_{0}k_{1}
+q_{1}k_{0}\right)\left(q_{0}+k_{0}\right)+q_{1}k_{1}^{2}+q_{1}^{2}k_{1}\right]\end{cases}\label{eq:second propds comp law}\end{equation}
Our proper-dS momentum space
is characterized (to quadratic order in $\ell$)
 by this commutative composition law
 and the on-shellness determined by the dS metric
on momentum space.

The composition law (\ref{eq:second propds comp law})
is not associative.
We opt to show this by first deriving some more general results.
Specifically we observe
that when composition laws are derived from the momentum-space affine connection
using our prescription, so that (\ref{eq:2 order comp law}) holds,
the associativity (or lack thereof) of the composition
law is linked to the connection, at second order in $\ell$, through
the relationship 
\begin{equation}
\left(\left(q\oplus_{\ell}k\right)\oplus_{\ell}p\right)_{\lambda}
-\left(q\oplus_{\ell}\left(k\oplus_{\ell}p\right)\right)_{\lambda}
=-\frac{\ell^2}{2}\left(\bar F^{\rho\beta\alpha}{}_{\lambda}
+\nabla_\Gamma^{\rho}\bar T^{\alpha\beta}{}_{\lambda}\right)k_{\alpha}p_{\beta}q_{\rho}
\label{joc26july}
\end{equation}
 where $\ell^{2} \nabla_\Gamma \bar T$ and $\ell^{2} \bar F$ are, respectively, the covariant derivative of the torsion contribution to the connection and 
the curvature of the connection, so that in particular
 \begin{equation}
\bar F^{\rho\beta\alpha}{}_{\lambda}=\partial^{\rho}\bar{\Gamma}^{\beta\alpha}{}_{\lambda}
-\partial^{\beta}\bar{\Gamma}^{\rho\alpha}{}_{\lambda}+\bar{\Gamma}^{\rho\tau}{}_{\lambda}\bar{\Gamma}^{\beta\alpha}{}_{\tau}
-\bar{\Gamma}^{\beta\tau}{}_{\lambda}\bar{\Gamma}^{\rho\alpha}{}_{\tau}
\label{joc26julyB}
\end{equation}
Again in (\ref{joc26july}) and  (\ref{joc26julyB})
it is intended that the fields are  to be evaluated in the origin of momentum
space.

For the specific case of our proper-dS momentum space we evidently have no torsion
 and the curvature of the connection is, to all orders,
 \begin{equation}
 F^{\rho\beta\alpha}{}_{\lambda}(p)=2\ell^{2}\delta_{0}^{[\rho}\delta_{1}^{\beta]}
\left(e^{2\ell p_{0}}\delta_{1}^{\alpha}\delta_{\lambda}^{0}
+\delta_{\lambda}^{1}\delta_{0}^{\alpha}\right)
\end{equation}
which translates into a non-associativity of the form
\begin{equation}
\left(\left(q\oplus_{\ell}k\right)\oplus_{\ell}p\right)_{\lambda}
-\left(q\oplus_{\ell}\left(k\oplus_{\ell}p\right)\right)_{\lambda}
=\frac{\ell^{2}}{2}\left(\delta_{\lambda}^{0} k_{1}+\delta_{\lambda}^{1}k_{0}\right)\left(p_{0}q_{1}-p_{1}q_{0}\right)\end{equation}

We also notice in closing that for the proper-dS composition law (\ref{eq:second propds comp law})
both the left loop inverse rule and the right loop inverse rule are not verified.

\section{DSR-relativistic properties of proper-dS and $\kappa$-dS momentum spaces\label{sec:DSR-relativistic-properties-of}}
As stressed already in Sec.I we are not just interested in dS momentum spaces with commutative composition laws:
we are also looking for rules of kinematics which are relativistic.
 Since special-relativistic laws of transformation cannot be symmetries
of any de Sitter momentum space,  relativistic
invariance must be inevitably described within the structure
of {}``DSR relativistic theories\textquotedbl{}~\cite{dsr1Edsr2}
(also see Refs.~\cite{jurekDSRfirst,leejoaoPRDdsr,gacDSRnature,leejoaoCQGrainbow,jurekDSRreview,gacSYMMETRYreview}),
theories with two relativistic invariants, the speed-of-light scale
$c$ and a length/inverse-momentum scale: the scale $\ell$ that characterizes
the geometry of momentum space must in fact be an invariant if the
theories on such momentum spaces are to be relativistic.

As already noticed in Ref. \cite{GACarXiv11105081} the requirement that a curved momentum
space be ``DSR compatible"
is strongly constraining: with arbitrary combinations of a metric
and of an affine connection one in general ends up with a combination of law of on-shellness
and law of energy-momentum conservation which allows the identification
of a preferred frame. But enforcing a suitable compatibility between
metric and affine connection one does find relativistic momentum spaces
\cite{GACarXiv11105081}, and we shall now show
that all 3 cases we considered, the $\kappa$-dS case, the proper-dS case and the weakly-proper-dS case,
are indeed DSR relativistic.

\subsection{The starting point of the relativistic properties
of Minkowski momentum space}

We set the stage for our analysis of relativistic properties of 1+1-dimensional
$\kappa$-dS and {}``proper-dS\textquotedbl{} momentum spaces by
first reviewing a few known facts about the special-relativistic properties
of the standard Minkowski momentum space.

For Minkowski momentum space of course the on-shellness has the form
\begin{equation}
m^{2}(p)=E^{2}-p^{2}=\eta^{\mu\nu}p_{\mu}p_{\nu}\end{equation}
 and the conservation of energy-momentum has the form
 \begin{equation}
p_{\mu}=q_{\mu}+k_{\mu}\end{equation}
 (focusing again on the example of a two-body particle decay).

The relativistic compatibility of Minkowski momentum space is ensured
by the standard special-relativistic rules of transformation of momenta,
which in particular for boosts we shall characterize in terms of
\begin{equation}
p_{\mu}\rightarrow\Lambda_{\mu}^{\xi}(p)\end{equation}
 where indeed we denoted by $\Lambda^{\xi}$ a standard Lorentz boost of
 rapidity/boost parameter $\xi$,
with generator $N$ such that (for small $\xi$)

\begin{equation}
p_{\mu}\rightarrow\Lambda_{\mu}^{\xi}(p)\approx p_{\mu}+\xi N_{\mu}(p)\label{eq:De Sitter boost}\end{equation}

Indeed under these standard boosts one easily finds that the on-shellness
is observer independent,
\begin{equation}
m^{2}(\Lambda^{\xi}(p))=m^{2}(p)~,\end{equation}
 and also the composition law is observer independent,
 \begin{equation}
p_{\mu}=q_{\mu}+k_{\mu}\rightarrow\Lambda_{\mu}^{\xi}(p)
=\Lambda_{\mu}^{\xi}(q)+\Lambda_{\mu}^{\xi}(k).\end{equation}

We also note briefly that in special relativity the on-shellness is
also invariant under momentum-space translations. To see in which
sense this holds let us consider two arbitrary points on the Minkowskian
momentum space, $k$ and $q$, and their distance
\begin{equation}
d^{2}(k,q)=\eta^{\mu\nu}\left(k_{\mu}-q_{\mu}\right)\left(k_{\mu}-q_{\mu}\right)\end{equation}
 This is not only invariant under boosts but also under the full Poincar\'e
group, with momentum-space translations of course given by
\begin{equation}
q_{\mu}\rightarrow q_{\mu}+a_{\mu}\end{equation}

\subsection{DSR-relativistic compatibility of {}``proper dS\textquotedbl{}
momentum space}

Let us now start considering the case of our {}``proper-dS\textquotedbl{}
momentum space. As we shall see this is a case where the new laws
of kinematics
\begin{equation}
\begin{cases}
m^{2}=d_{\ell}^{2}(p)\\
p_{\mu}=\left(q\oplus_{\ell}k\right)_{\mu}\end{cases}\end{equation}
 are still compatible with a rather standard implementation of observer
independence, involving of course DSR-deformed boost transformations
${\tilde{\Lambda}}$. As we shall soon show, the action of these deformed boosts on a momentum $p_\mu$ is governed by
\begin{equation}
B^{\xi}\triangleright p_\mu ={\tilde{\Lambda}}^{\xi}_\mu(p)
\approx p_{\mu}+\xi {\tilde{N}}_{\mu}(p)
\label{jocNN1}\end{equation}
with
\begin{equation}
\begin{cases}
{\tilde{N}}_{0}(p)=p_{1}\\
{\tilde{N}}_{1}(p)
=\frac{1}{2\ell}\left(1-e^{-2\ell p_{0}}\right)-\frac{\ell}{2}p_{1}^{2}\end{cases}
\label{De Sitter boost2}\end{equation}

It is easy to check (see below) that the
condition of on-shellness is observer independent
under these transformation laws
\begin{equation}
d_\ell^{2}\left(B^{\xi}\triangleright p\right)=d_\ell^{2}\left(p\right)=m_{p}^{2}\end{equation}

It is already well
established~\cite{GACarXiv11105081,dsr1Edsr2,jurekDSRreview,gacSYMMETRYreview}
that the full (DSR-)relativistic
compatibility of a momentum space also imposes strong demands on
the action of boosts on
 composed momenta: $B_{\xi}\triangleright\left(a\oplus_{\ell} b\right)$.
This is the part that is most challenging from a relativistic perspective
and it shall prove to be particularly striking when we get to the point
of analyzing the relativistic properties of the $\kappa$-momentum space.
For the case of our proper-dS momentum space
 it turns out that the action of boosts on composed
momenta retains most of its intuitive properties, and in particular is such that
\begin{equation}
B^{\xi}\triangleright\left(q\oplus_{\ell}k\right)={\tilde{\Lambda}}^{\xi}(q)\oplus_{\ell}{\tilde{\Lambda}}^{\xi}(k)\label{masterDSPROP}\end{equation}
with
\begin{equation}\label{eq:standard cov of comp}
B^{\xi}\triangleright\left(q\oplus_{\ell}k\right) = {\tilde{\Lambda}}_{\mu}^{\xi}({\cal P})\Big|_{{\cal P}=q\oplus_{\ell}k}\end{equation}
All this can be summarized,
in the spirit and notation of Ref.\ttt \cite{GACarXiv11105081}, in terms of the following
intuitive property of boost generators:
 \begin{equation}
{\tilde{N}}^{\xi}_{[q \oplus_{\ell} k]}
= {\tilde{N}}^{\xi}_{[q]} + {\tilde{N}}^{\xi}_{[k]}
\end{equation}
meaning that the action of the boost generator on the composed momenta is simply obtained
by acting on each momentum in the composition, following a standard implementation of Leibniz rule.

We shall provide support for these claims  while observing incidentally that
 also for our {}``proper dS\textquotedbl{} momentum space (just
like stressed at the end of the previous subsection for the standard
case of Minkowskian momentum space)  the distance
between points of momentum space is invariant not only under boosts
but also under some appropriate notion of momentum-space translations.
So we look for transformations ${\cal S}$ that leave invariant
the distance on momentum space: \begin{equation}
d_{\ell}^{2}({\cal S}(k),{\cal S}(q))=d_{\ell}^{2}(k,q)\label{eq:invariance of distance}\end{equation}
where ${\cal S}$ must be regarded as a particular class
of diffeomorphisms on momentum space:
\begin{equation}
{\cal S}:p_{\mu}\rightarrow{\cal S}_{\mu}(p)\label{eq:S is a diffeo}\end{equation}
which solves (\ref{eq:invariance of distance}) and reduces to the
undeformed Poincar\'e group of transformations in the $\ell\rightarrow0$ limit.

We focus on the infinitesimal
set of diffeomorphisms:
\begin{equation}
{\cal S}_{\mu}(p)\approx p_{\mu}+{\cal T}_{\mu}(p)\end{equation}
where the transformations ${\cal T}$ can be considered as
the generators
of the deformed Poincar\'e group. In terms of the generators ${\cal T}$
then the request of invariance of the distance on momentum space takes
the form:
\begin{equation}
d_{\ell}^{2}(k+{\cal T}(k),q+{\cal T}(q))=d_{\ell}^{2}(k,q)\end{equation}
which we are going to solve at first order in ${\cal T}$.

By defining $L(\gamma,\dot{\gamma})=\sqrt{g^{\mu\nu}(\gamma)\dot{\gamma}_{\mu}\dot{\gamma}_{\nu}}$
and using the definition of $d_\ell^{2}$ one easily finds
\begin{equation}\label{eq:variation of distance5}
\begin{array}{c}
0=\delta\int dtL(\gamma(t),\dot{\gamma}(t))=\int\left(\frac{\partial L}{\partial\gamma}-\frac{d}{dt}\left(\frac{\partial L}{\partial\dot{\gamma}}\right)\right)\delta\gamma dt+\left(\frac{\partial L}{\partial\dot{\gamma}}\delta\gamma\right)|_{t=0}^{t=1}\end{array}\end{equation}
 where $\gamma$ is the curve connecting $q$ to $k$ and $\delta\gamma(t)$
is the variation of $\gamma(t)$ due to the variation of its extremal
points $q$ and $k$, which must satisfy $\delta\gamma(0)={\cal T}(q)$
and $\delta\gamma(1)={\cal T}(k)$ .

Using the definition of $L$ the integrand term on the right-hand side of (\ref{eq:variation of distance5}) is easily seen to be the geodesic equation which is solved, by definition, by the curve $\gamma$.
We are then left with the boundary term:
 \begin{equation}
\frac{d}{dt}\left(\frac{\partial L}{\partial\dot{\gamma}}\delta\gamma\right)|_{t=0}^{t=1}=0\end{equation}

We show in appendix (\ref{sec:cov of distance app})
that, enforcing the validity of the previous equation for any couple of points $q$ and $k$,  translates in
a requirement for the generators ${\cal T}$ of the form \begin{equation}
\nabla_{(A)}^{(\mu}{\cal T}^{\nu)}=0\label{jocnabla}\end{equation}
 where $\nabla_{A}$ is the covariant derivative associated to the
Levi-Civita connection $A^{\lambda\mu}{}_{\nu}$ defined by the momentum-space
metric $g^{\mu\nu}$. Evidently we found, with (\ref{jocnabla}),
that ${\cal T}$ must satisfy a killing equation. And this killing
equation admits solution for three classes of metrics, the ones that
are diffeomorphic to Minkowski, de Sitter or anti de Sitter metric.
We are here focusing for definiteness on the de Sitter case.

For the de Sitter momentum space, still working in 1+1 dimension,
it is easy to show, as we do in appendix \ref{sec:cov of distance app},
that the solutions of (\ref{jocnabla}) are: \begin{equation}
\begin{cases}
{\cal T}_{0}(p)=\xi p_{1}+\gamma\\
{\cal T}_{1}(p)=\xi\left(\frac{1}{2\ell}\left(1-e^{-2\ell p_{0}}\right)-\frac{\ell}{2}p_{1}^{2}\right)-\ell p_{1}\gamma+\beta\end{cases}\label{eq:De Sitter symm}\end{equation}
 Here, $\xi$ , $\beta$ and $\gamma$ arise as constants of integration
of the killing equation, and have to be taken as the infinitesimal
parameters associated to the boost and (two) translations in the de
Sitter momentum space.

The $\ell$-deformed boost generators, which are of our primary interest,
can be easily read off (\ref{eq:De Sitter symm}): \begin{equation}
\begin{cases}
{\tilde{ N}}_0(p)=p_{1}\\
{\tilde{N}}_{1}(p)=\left(\frac{1}{2\ell}\left(1-e^{-2\ell p_{0}}\right)-\frac{\ell}{2}p_{1}^{2}\right)\end{cases}
\end{equation}
which, confirming our earlier claims, coincide with (\ref{De Sitter boost2}).

This choice of deformed boost transformations ensures by construction
the observer independence of the on-shell relation. Our next task
is to show that it also ensures the observer independence of
energy-momentum conservation laws, which we shall investigate at the level of the
properties of the
the law $\oplus_\ell$ of composition of momenta. In this subsection we shall be satisfied
performing this analysis at second
order in $\ell$.
 We start by noticing that at second
order in $\ell$ the boost generator (\ref{De Sitter boost2}) takes
the form \begin{equation}
\begin{cases}
{\tilde{ N}}_0(p) \simeq p_{1}\\
{\tilde{ N}}_1(p) \simeq \left(p_{0}-\ell p_{0}^{2}-\frac{\ell}{2}p_{1}^{2}
+\frac{1}{3}\ell^{2}p_{0}^{3}\right)\end{cases}
\label{eq:ds boost}\end{equation}

The announced DSR-relativistic compatibility of our proper-dS momentum space,
taking the form of (\ref{masterDSPROP})-(\ref{eq:standard cov of comp})
is automatically ensured by verifying that (at leading order in the boost parameter $\xi$)
\begin{equation}
{\tilde{ N}}_{\mu}\left({\cal P}\right)\Big|_{{\cal P}=q\oplus_{\ell}k}
= \frac{\partial}{\partial q_{\rho}}\left(q\oplus_{\ell}k\right)_{\mu}
{\tilde{N}}_{\rho}(q) +\frac{\partial}{\partial k_{\rho}}\left(q\oplus_{\ell}k\right)_{\mu}{\tilde {N}}_{\rho}(k)
\label{joczwei}\end{equation}
 The validity of this (\ref{joczwei}) is easily checked by computing separately its two sides and finding that they give the
same result. It suffices  to substitute the explicit form of the second
order boost (\ref{eq:ds boost}) and of the second order proper-dS
composition law computed in (\ref{eq:second propds comp law}),
so that one gets for the left-hand side of (\ref{joczwei})
\begin{equation}
{\tilde{ N}}_{0}\left({\cal P}\right)\Big|_{{\cal P}=q\oplus_{\ell}k}=
q_{1}+k_{1}-\ell\left(q_{0}k_{1}+q_{1}k_{0}\right)+\frac{\ell^{2}}{2}\left[\left(q_{0}k_{1}+q_{1}k_{0}\right)\left(q_{0}
+k_{0}\right)+q_{1}k_{1}^{2}+q_{1}^{2}k_{1}\right]
\label{eq:prop comp is rel 1}\end{equation}
and
\begin{equation}
\begin{array}{c}
{\tilde{ N}}_{1}\left({\cal P}\right)\Big|_{{\cal P}=q\oplus_{\ell}k}=q_{0}+k_{0}-l\left[q_{1}k_{1}+\left(q_{0}+k_{0}\right)^{2}+\frac{1}{2}\left(q_{1}+k_{1}\right)^{2}\right]+l^{2}\left(\frac{5}{2}q_{0}q_{1}k_{1}+\frac{5}{2}k_{0}q_{1}k_{1}+\frac{3}{2}q_{1}^{2}k_{0}+\frac{3}{2}q_{0}k_{1}^{2}\right)+\frac{2}{3}l^{2}\left(q+k\right)_{0}^{3}\end{array}\label{eq:prop comp is rel 4}\end{equation}
while for the right-hand side
of (\ref{joczwei}) one gets
\begin{equation}
\begin{array}{c}
\frac{\partial}{\partial q_{\rho}}\left((q\oplus_{\ell}k)_{0}\right)
{\tilde{ N}}_{\rho}(q)+\frac{\partial}{\partial k_{\rho}}\left((q\oplus_{\ell}k)_{0}\right)
{\tilde{ N}}_{\rho}(k)=\\
=\frac{\partial}{\partial q_{\rho}}\left(q_{0}+k_{0}-\ell q_{1}k_{1}+\frac{\ell^{2}}{2}\left[-q_{1}k_{1}\left(q_{0}+k_{0}\right)+q_{0}k_{1}^{2}+q_{1}^{2}k_{0}\right]\right)
{\tilde{ N}}_{\rho}(q)+\\
+\frac{\partial}{\partial k_{\rho}}\left(q_{0}+k_{0}-\ell q_{1}k_{1}
+\frac{\ell^{2}}{2}\left[-q_{1}k_{1}\left(q_{0}+k_{0}\right)+q_{0}k_{1}^{2}+q_{1}^{2}k_{0}\right]\right)
 {\tilde{ N}}_{\rho}(k)=\\
=q_{1}+k_{1}-l\left(k_{1}q_{0}+q_{1}k_{0}\right)+\frac{l^{2}}{2}\left[k_{1}q_{0}^{2}+k_{1}q_{1}^{2}+q_{1}k_{0}^{2}+q_{1}k_{1}^{2}+q_{0}k_{1}k_{0}+q_{1}q_{0}k_{0}\right]\end{array}\label{eq:prop comp is rel 2}\end{equation}
and
\begin{equation}
\begin{array}{c}
\frac{\partial}{\partial q_{\rho}}\left((q\oplus_{\ell}k)_{1}\right)
{\tilde{ N}}_{\rho}(q)+\frac{\partial}{\partial k_{\rho}}\left((q\oplus_{\ell}k)_{1}\right)
{\tilde{ N}}_{\rho}(k)=\\
=\frac{\partial}{\partial q_{\rho}}\left(q_{1}+k_{1}-\ell\left(q_{0}k_{1}+q_{1}k_{0}\right)+\frac{\ell^{2}}{2}\left[\left(q_{0}k_{1}+q_{1}k_{0}\right)\left(q_{0}+k_{0}\right)+q_{1}k_{1}^{2}+q_{1}^{2}k_{1}\right]\right)
{\tilde{ N}}_{\rho}(q)+\\
+\frac{\partial}{\partial k_{\rho}}\left(q_{1}+
k_{1}-\ell\left(q_{0}k_{1}+q_{1}k_{0}\right)
+\frac{\ell^{2}}{2}\left[\left(q_{0}k_{1}+q_{1}k_{0}\right)\left(q_{0}
+k_{0}\right)+q_{1}k_{1}^{2}+q_{1}^{2}k_{1}\right]\right)
{\tilde{ N}}_{\rho}(k)=\\
=q_{0}+k_{0}-l\left(q_{1}k_{1}+\left(q_{0}+k_{0}\right)^{2}+\frac{1}{2}\left(q_{1}+k_{1}\right)^{2}\right)
+l^{2}\left(\frac{5}{2}q_{0}q_{1}k_{1}+\frac{5}{2}k_{0}q_{1}k_{1}+\frac{3}{2}q_{1}^{2}k_{0}+\frac{3}{2}q_{0}k_{1}^{2}\right)+\frac{2}{3}l^{2}\left(q+k\right)_{0}^{3}\end{array}\label{eq:prop comp is rel 3}\end{equation}

This completes our verification of the (DSR-)relativistic compatibility
of the {}``proper dS\textquotedbl{} momentum space to second order
in $\ell$.\\
 We stress that by checking the observer independence of the law of composition of two momenta we automatically also ensured the observer independence
of the composition of any number of momenta. For example, one easily sees that
 \begin{equation}
{\tilde{\Lambda}}^{\xi}_\mu({\cal P})\Big|_{{\cal P}=a\oplus_{\ell}b}
={\tilde{\Lambda}}^{\xi}_\mu(a)\oplus_\ell {\tilde{\Lambda}}^{\xi}_\mu(b)~~ \Longrightarrow ~~
{\tilde{\Lambda}}^{\xi}_\mu({\cal P})\Big|_{{\cal P}=a\oplus_{\ell}b\oplus_{\ell}c}=
{\tilde{\Lambda}}^{\xi}_\mu({\cal P}')\Big|_{{\cal P}'=a\oplus_{\ell}b}\oplus_\ell {\tilde{\Lambda}}^{\xi}_\mu(c)=
{\tilde{\Lambda}}^{\xi}_\mu(a)\oplus_\ell {\tilde{\Lambda}}^{\xi}_\mu(b)
\oplus_\ell {\tilde{\Lambda}}^{\xi}_\mu(c)
\label{jocmulti}\end{equation}

\subsection{DSR compatibility of weakly-proper de Sitter\label{sec:DSR compatibility of Improper de Sitter}}
Let us now comment on the DSR-relativistic compatibility of the ``weakly-proper dS momentum space" which we introduced in
Subsection \ref{sec:Improper dS}. This is the scenario we obtained adopting on momentum space de Sitter metric and
 its Levi-Civita connection, but then requiring that the composition law be commutative and compatible with the connection
  according to standard interpretation (\ref{eq: Connection Introduction}). Following the strategy of analysis already discussed
  in the previous subsection it is easy to see that also this weakly-proper dS momentum space is DSR-relativistic compatible.
And the core ingredient of this relativistic compatibility is exactly the same as for the case of the proper-dS momentum space
discussed in the previous subsection. Indeed both scenarios have the same on-shellness (same metric on momentum space, the de Sitter one)
whose invariance is assured by (at second order in $\ell$ and at first order in the boost parameter
$\xi$)
\begin{equation}
\begin{cases}
\tilde{\Lambda}_{0}(p)\simeq p_{0}+\xi p_{1}\\
\tilde{\Lambda}_{1}(p)\simeq p_{1}+\xi\left(p_{0}-\ell p_{0}^{2}-\frac{\ell}{2}p_{1}^{2}+\frac{2}{3}\ell^{2}p_{0}^{3}\right)\end{cases}\label{eq:improp boost}\end{equation}
The differences between the weakly-proper-dS and the proper-dS case all reside in the law of composition of momenta, which in the
weakly-proper-dS case takes the form
\begin{equation}\label{eq:improper comp law}
\begin{cases}
\left(q\oplus_{\ell}k\right)_{0}^{(2)}=q_{0}+k_{0}-\ell q_{1}k_{1}\\
\left(q\oplus_{\ell}k\right)_{1}^{(2)}=q_{1}+k_{1}-\ell\left(q_{0}k_{1}
+q_{1}k_{0}\right)+\ell^{2}k_{1}\left(q_{0}^{2}+\frac{1}{2}q_{1}^{2}\right)+\ell^{2}q_{1}\left(k_{0}^{2}+\frac{1}{2}k_{1}^{2}\right)
\end{cases}\end{equation}
It is noteworthy that, in spite of their differences, both this composition law for the weakly-proper-dS case and the composition law for the
proper-dS are covariant under the boosts (\ref{eq:improp boost}).
We leave to the interested reader the simple task of verifying that indeed for the weakly-proper composition law (\ref{eq:improper comp law})
one does have that
\begin{equation}
\tilde{\Lambda}^{\xi}\left(q\oplus_{\ell} k\right)=\tilde{\Lambda}^{\xi}(q)\oplus_{\ell}\tilde{\Lambda}^{\xi}(k)\label{eq:cov of impr comp}\end{equation}
So we have more than one (at least two\footnote{We here explicitly obtained two commutative composition laws (proper-dS case and weakly-proper-dS case) which satisfy this requirement of DSR-relativistic compatibility with the on-shellness obtained from the dS metric of momentum space.
It goes beyond the scopes of this analysis to determine how many such commutative composition laws exist. Concerning noncommutative composition
laws we here consider the single case of the $\kappa$-momentum space, with its noncommutative composition law which
is also (see subsection \ref{subsec:DSRkappa}) DSR-relativistic compatible with the on-shellness obtained from the dS metric of momentum space.
And as we were in the final stages of preparation of this manuscript
we noticed the very recent study in Ref. \cite{Banburski:2013wxa} which provides another example of
noncommutative composition law which
is DSR-relativistic compatible with the on-shellness obtained from the dS metric of momentum space.}
commutative composition laws
which can be consistently used in a DSR-relativistic picture
in combination with the on-shellness obtained from the dS metric of momentum space.

\subsection{Aside on more general forms of DSR-relativistic compatibility}
Both for the proper-dS and for the weakly-proper-dS momentum spaces we established a
form of DSR-relativistic compatibility which is still rather intuitive.
This is based on requiring for the action of boosts on composed momenta that
$$B^{\xi}\triangleright\left(q\oplus_{\ell}k\right)={\tilde{\Lambda}}^{\xi}(q)\oplus_{\ell}{\tilde{\Lambda}}^{\xi}(k)$$
with
$$B^{\xi}\triangleright\left(q\oplus_{\ell}k\right) = {\tilde{\Lambda}}_{\mu}^{\xi}({\cal P})\Big|_{{\cal P}=q\oplus_{\ell}k}$$
where $p \rightarrow {\tilde{\Lambda}}^{\xi}(p)$ is a DSR-deformed ($\ell$-dependent) boost map which is compatible
with the momentum-space metric in the sense that it leaves the on-shellness invariant.

And this intuitive setup also allows one to describe the action of boosts on composed momenta
via generators of the form
 \begin{equation}
{\tilde{N}}^{\xi}_{[q \oplus_{\ell} k]}
= {\tilde{N}}^{\xi}_{[q]} + {\tilde{N}}^{\xi}_{[k]}
\label{compboostDSPROP}
\end{equation}
meaning that the action of the boost generator on the composed momenta is simply obtained
by acting on each momentum in the composition with the single-momentum generator (the operator which generates
 the boost
map $p \rightarrow {\tilde{\Lambda}}^{\xi}(p)$).

Our next task is to observe that this particularly simple structure of the DSR-relativistic properties
is not representative of the most general case. One can achieve the equivalence of inertial observers (``absence
of a preferred frame") also in some less intuitive and more complicated ways. We shall dwell on this not only for its
intrinsic interest from a general perspective but also because the DSR-relativistic compatibility  of the $\kappa$-momentum
space can only be understood from within this wider picture, as we shall show in the next subsection.

We get on our way toward discussing these issues by assuming that it has been already established that
some description of the boost map on the momentum of a single particle,
\begin{equation}
B^{\xi}\triangleright p_\mu ={\tilde{\Lambda}}^{\xi}_\mu(p)
\approx p_{\mu}+\xi {\tilde{N}}_{\mu}(p) \,  ,
\label{jocNN1b}\end{equation}
is relativistically compatible with the on-shellness
\begin{equation}
d_\ell^{2}\left(B^{\xi}\triangleright p\right)=d_\ell^{2}\left(p\right)=m_{p}^{2}\end{equation}

The part which can be highly non-trivial~\cite{GACarXiv11105081,dsr1Edsr2,jurekDSRreview,gacSYMMETRYreview}
is the formulation of DSR-relativistic properties for the law of composition of momenta, {\it i.e.}
concerns
the action of boosts on
 composed momenta: $B_{\xi}\triangleright\left(q \oplus_\ell k\right)$.
 It should be observed that one obtains a DSR-relativistic picture if
 the following  properties are established:\\
$\bullet$ A first requirement is that the law of composition of momenta be \lqt covariant in substance", by which we mean that
 the action of boosts on composed momenta should be such that conservation laws are observer independent.
  For the case of a two-body particle decay this takes the shape of the requirement
 \begin{equation}
p=q\oplus_{\ell}k\Rightarrow B^{\xi}\triangleright p=B^{\xi}\triangleright\left(q\oplus_{\ell}k\right)\label{eq:conservation of conservation lawb}\end{equation}
Evidently this requirement is satisfied if the action of boosts on a momentum obtained combining two other momenta
follows the same law of transformation of a corresponding single momentum:
\begin{equation}
B^{\xi}\triangleright\left(q\oplus_{\ell}k\right)
={\tilde{\Lambda}}^{\xi}_\mu({\cal P})\Big|_{{\cal P}=q\oplus_{\ell}k}
\label{jocsubstb}
\end{equation}
\noindent
$\bullet$ A second requirement
is that the law of composition of momenta be \lqt covariant in form", by which we mean that
 the action of boosts on momenta composed via the $\oplus_\ell$ rule produces momenta which are still composed
 via the $\oplus_\ell$ rule.
 We make explicit this requirement by introducing notation $p^{[A]}$,$q^{[A]}$ for the momenta of two particles
 measured by Alice as part of a composition law (momenta of particles taking part in an event, and therefore entering a
  law of conservation of energy-momentum) and notation $p^{[B]}$,$q^{[B]}$ for the momenta of those same two
  particles as determined by observer Bob, purely boosted with respect to Alice.
  The requirement then takes the form of
\begin{equation}
B^{\xi}\triangleright\left(q^{[A]} \oplus_{\ell} k^{[A]} \right)
=q^{[B]} \oplus_{\ell} k^{[B]}
\label{jocformb}
\end{equation}
$\bullet$ As third requirement we shall insist on having that $q^{[B]}$ (respectively $k^{[B]}$) is on the same shell
as $q^{[A]}$ (respectively $k^{[A]}$).

The first requirement ensures that processes which are allowed for observer Alice ({\it i.e.} such that momentum
is indeed conserved for Alice) are also allowed for Bob ({\it i.e.} momentum
is conserved also according to Bob). Moreover, Alice and Bob agree (because of the second requirement) on the form of the composition law
 and also agree on which particles take part in the process (third requirement).
So there is no way to distinguish between Alice and Bob: the laws of kinematics are exactly the same for Alice and Bob, as required
for a relativistic picture.

The intuitive case of DSR-relativistic pictures which we encountered for the proper-dS and weakly-proper-dS cases
satisfies these requirements in a particularly simple way: the boost $B^{\xi}$ of rapidity $\xi$ acts on composed momenta in
such a way that $q^{[B]}$ is just ${\tilde{\Lambda}}^{\xi}_\mu(q^{[A]})$ and $k^{[B]}$ is just ${\tilde{\Lambda}}^{\xi}_\mu(k^{[A]})$.
This evidently ensures that, as dictated by the third requirement, $q^{[B]}$ (respectively $k^{[B]}$) is on the same shell
as $q^{[A]}$ (respectively $k^{[A]}$). But there are more general ways to ensure the third  requirement, while not
spoiling the first and second requirement. In particular,
one can allow for $q^{[B]}$ to be ${\tilde{\Lambda}}^{\xi_1}_\mu(q^{[A]})$ and $k^{[B]}$ to be ${\tilde{\Lambda}}^{\xi_2}_\mu(k^{[A]})$,
with any choice of $\xi_1$ and $\xi_2$, and still the third requirement is satisfied:
\begin{equation}
B^{\xi}\triangleright\left(q^{[A]} \oplus_{\ell} k^{[A]} \right)
={\tilde{\Lambda}}^{\xi_{1}}(q^{[A]})\oplus_{\ell}{\tilde{\Lambda}}^{\xi_{2}}(k^{[A]})
\label{jocformc}
\end{equation}
The $\kappa$-momentum space will give us
an explicit example where this is the way in which the third requirement is satisfied, preserving the compatibility
with the first and second requirement. As we shall see is  the case of $\kappa$-momentum space, one in general expects $\xi_1$
and $\xi_2$ to be proportional to the rapidity parameter $\xi$ of the boost, but having $\xi_1 = \xi = \xi_2$ is not the only
admissible way (for the $\kappa$-momentum space we shall find that $\xi_1 = \xi$ but $\xi_2 \neq \xi$). Of course any difference
between $\xi_1$ and $\xi$ and/or between $\xi_2$ and $\xi$ must be governed by the only relevant physical
observables, which are the momenta which are being composed.

In closing this aside we also notice that (\ref{jocformc}) can be rewritten for generators
in the following way
 \begin{equation}
{\tilde{N}}^{\xi}_{[q \oplus_{\ell} k]}
= {\tilde{N}}^{\xi_1}_{[q]} + {\tilde{N}}^{\xi_2}_{[k]}
\label{compboostDSPROPb}\end{equation}
a special case of which is (\ref{compboostDSPROP}).

\subsection{DSR-relativistic compatibility of $\kappa$-dS momentum space}
\label{subsec:DSRkappa}
We are now ready for the task of establishing the DSR-relativistic compatibility
of the $\kappa$-dS momentum space. We first note that the action
of a Lorentz boost on single momenta $ {\tilde{\Lambda}}^{\xi}_\mu(p)$
is the same
we encountered for the ``proper-dS'' and ``weakly-proper-dS'' momentum spaces:
\begin{equation}
\begin{cases}
\tilde N_{0}(p)=p_{1}\\
\tilde N_{1}(p_{0})=\left(\frac{1}{2\ell}\left(1-e^{-2\ell p_{0}}\right)-\frac{\ell}{2}p_{1}^{2}\right)\end{cases}\label{kappaboost}\end{equation}
This is due to the fact that also on the $\kappa$-dS momentum space the metric
is the de Sitter metric (so the action of boosts on single momenta must
enforce the observer independence of the same on-shellness). The differences
are all at the level of the composition law, and particularly significant
for our purposes is the fact that on the
$\kappa$-dS momentum space the composition law is noncommutative:
\begin{equation}
\begin{cases}
\left(q\oplus_{\ell}k\right)_{0}=q_{0}+k_{0}\\
\left(q\oplus_{\ell}k\right)_{1}=q_{1}+k_{1}e^{-\ell q_{0}}\end{cases}
\label{eq:kappa comp law}\end{equation}
This noncommutativity is evidently a challenge~\cite{GACarXiv11105081} for a symmetric rule of action of boosts on composed momenta, of the
type $B^{\xi}\triangleright\left(q \oplus_{\ell} k \right) =
{\tilde{\Lambda}}^{\xi}(q)\oplus_{\ell}{\tilde{\Lambda}}^{\xi}(k)$. We shall see that the $\kappa$-dS momentum space is DSR-relativistic compatible
but only according to the sort of relativistic prescriptions described in the previous subsection,  such that
\begin{equation}
B^{\xi}\triangleright\left(q \oplus_{\ell} k \right) \neq
{\tilde{\Lambda}}^{\xi}(q)\oplus_{\ell}{\tilde{\Lambda}}^{\xi}(k)
\label{jocNOTsoEASY}\end{equation}
Building on results of previous studies of the $\kappa$-dS momentum space~\cite{GiuliaFlavio, GACarXiv11105081, GACarxiv1111.5643}
we shall give a more detailed account than
ever before of the relativistic properties of the $\kappa$-dS momentum space,
when adopting as rule of action on the composition of two momenta
the following \cite{GiuliaFlavio}
\begin{equation}
B^{\xi}\triangleright\left(q \oplus_{\ell} k\right) =
{\tilde{\Lambda}}^{\xi}(q)
\oplus_{\ell}{\tilde{\Lambda}}^{\xi e^{-\ell q_{0}}}(k)
\label{jocNOTsoEASYbis}\end{equation}
where we are here satisfied with working
 at leading order\footnote{A generalization of (\ref{jocNOTsoEASYbis})
 to all orders in $\xi$ can be found in
 Ref.~\cite{GiuliaFlavio}.} in $\xi$ and
the symbol ${\tilde{\Lambda}}^{\xi e^{-\ell q_{0}}}$
is intended as just
one of our deformed boosts but with the peculiarity
that the rapidity parameter is specified not only in terms of the rapidity parameter connecting the two relevant observers
but also in terms of the (exponential of the zeroth component of) the momentum
appearing on the left-hand side of the composition law.

It is important to notice that (\ref{jocNOTsoEASYbis})
automatically enforces the ``covariance in form" characterized,
as stressed earlier,  by the requirement
$B^{\xi}\triangleright\left(q^{[A]} \oplus_{\ell} k^{[A]} \right)
=q^{[B]} \oplus_{\ell} k^{[B]}$: for this one should
 interpret (\ref{jocNOTsoEASYbis})
as establishing that the momenta determined by the two observers, Alice and Bob, connected
 by the boosts are related according to $q^{[B]} = {\tilde{\Lambda}}^{\xi}(q^{[A]})$
and $k^{[B]} =   {\tilde{\Lambda}}^{\xi e^{-\ell q_{0}}}( k^{[A]})$.
Notice that adopting this interpretation one still has (since the only actions are via ${\tilde{\Lambda}}$,
which by construction leaves the shell unchanged)
that $k^{[B]}$ and $k^{[A]}$ (and respectively $q^{[B]}$ and $q^{[A]}$)
are on the same mass shell, as necessary for relativistic invariance of the theory.
Our remaining task is to verify that our formulation of relativistic
transformations of $\kappa$-dS composed momenta is \lqt covariant in substance", in the
sense specified earlier through the requirement
$B^{\xi}\triangleright\left(a\oplus_{\ell}b\right)
={\tilde{\Lambda}}^{\xi}_\mu({\cal P})\Big|_{{\cal P}=a\oplus_{\ell}b}
$ or equivalently in the $\kappa$-dS case (also taking in to account (\ref{jocNOTsoEASYbis}))
the requirement~\cite{GACarXiv11105081,GACarxiv1111.5643}
\begin{equation}
\xi {\tilde{ N}}_{\mu}\left({\cal P}\right)\Big|_{{\cal P}=q\oplus_{\ell}k}
=\xi \frac{\partial}{\partial q_{\rho}}\left(q\oplus_{\ell}k\right)_{\mu}{\tilde{N}}_{\rho}(q)
 +\xi e^{-\ell q_{0}}\frac{\partial}{\partial k_{\rho}}\left(q\oplus_{\ell}k\right)_{\lambda}{\tilde {N}}_{\rho}(k)
 \label{eq:kappa inf covariance}\end{equation}

Before verifying this (\ref{eq:kappa inf covariance}) let us pause briefly for observing
that Eq.\ttt(\ref{jocNOTsoEASYbis})
and Eq.\ttt(\ref{eq:kappa inf covariance}) provide intuition for
the nature of the peculiarities of the $\kappa$-dS momentum space:
Eq.\ttt(\ref{jocNOTsoEASYbis}) lends itself to the perspective of how boosts act on each of the momenta
in the composition law and shows that for the second of the two momenta the boost parameter is affected
by a \lqt back reaction" \cite{GiuliaFlavio}
such that the first momentum in the composition law enters in the rapidity parameter \lqt felt by" the
second momentum in the composition law; instead Eq.\ttt(\ref{eq:kappa inf covariance})
 lends itself to the perspective of how the action of boost generators on composed momenta
 is structured
 in terms of the action of boosts generators on single momenta
 (in this case a rule of action of the type ${\tilde{N}}^{\xi}_{[q \oplus_{\ell} k]}
= {\tilde{N}}^{\xi}_{[q]} + e^{-\ell q_{0}} {\tilde{N}}^{\xi}_{[k]}$)
that leads to the notion \cite{GACarXiv11105081, GACarxiv1111.5643} of a \lqt law of composition of boosts".

Turning back to the \lqt covariance in substance" of this $\kappa$-dS case, we opt for
structuring our proof in such a way that we can expose the uniqueness
of the option (\ref{eq:kappa inf covariance}). We do so by considering
at first the slightly more general requirement
\begin{equation}
\xi {\tilde{ N}}_{\mu}\left({\cal P}\right)\Big|_{{\cal P}=q\oplus_{\ell}k}
=\xi_{(1)} \frac{\partial}{\partial q_{\rho}}\left(q\oplus_{\ell}k\right)_{\mu}{\tilde{N}}_{\rho}(q)
 +\xi_{(2)}\frac{\partial}{\partial k_{\rho}}\left(q\oplus_{\ell}k\right)_{\lambda}{\tilde {N}}_{\rho}(k)
\label{joctwoxi}\end{equation}
so that we can find as a result of the analysis that this requirement can
only be satisfied with the choice of $\xi_{(1)}$ and $\xi_{(2)}$ made in (\ref{eq:kappa inf covariance}),
{\it i.e.} $\xi_{(1)} = \xi$ and $\xi_{(2)} =  e^{-\ell q_{0}} \xi$.

For the zeroth component of (\ref{joctwoxi}), using the form of our ${\tilde{ N}}$
one easily finds that:
\begin{equation}
\begin{cases}
\xi {\tilde{ N}}_{0}\left({\cal P}\right)\Big|_{{\cal P}=q\oplus_{\ell}k}=\xi\left(q\oplus_{\ell}k\right)_{1}=\xi\left(q_{1}+k_{1}e^{-\ell q_{0}}\right)\\
\xi_{(1)}\frac{\partial}{\partial q_{\rho}}\left(q\oplus_{\ell}k\right)_{0}\tilde N_{\rho}(q)+\xi_{(2)}\left(\xi,q\right)\frac{\partial}{\partial k_{\rho}}\left(q\oplus_{\ell}k\right)_{0}\tilde N_{\rho}(k)=\xi_{(1)}q_{1}+\xi_{(2)}k_{1}\end{cases}
\end{equation}
From this one sees that the requirement
\begin{equation}\left(\xi-\xi_{(1)}\right)q_{1}+\left(\xi e^{-\ell q_{0}}-\xi_{(2)}\right)k_{1}=0\end{equation}
must be satisfied, which in turn
leads us to the sought conclusion $\xi_{(1)}=\xi$, $\xi_{(2)}=\xi e^{-\ell q_{0}}$
as the only option for having (\ref{joctwoxi}) verified for arbitrary $q$ and $k$.

Let us then reinstate $\xi_{(1)}=\xi$, $\xi_{(2)}=\xi e^{-\ell q_{0}}$ in (\ref{joctwoxi})
and verify that also the $1$ component of (\ref{joctwoxi}) is satisfied.
For this we observe that on the left-hand side of (\ref{joctwoxi}) we have
\begin{equation}
\begin{array}{c}
\tilde{N}_{1}\left({\cal P}\right)\Big|_{{\cal P}=q\oplus_{\ell} k}=\left(\frac{1}{2\ell}\left(1-e^{-2l\ell P_{0}}\right)-\frac{\ell}{2}{\cal P}_{1}^{2}\right)\Big|_{{\cal P}=q\oplus_{\ell} k}=\frac{1}{2\ell}\left(1-e^{-2\ell\left(q_{0}+k_{0}\right)}\right)-\frac{\ell}{2}\left(q_{1}+k_{1}e^{-\ell q_{0}}\right)^{2}=\\
=\frac{1}{2\ell}\left(1-e^{-2\ell\left(q_{0}+k_{0}\right)}\right)-\frac{\ell}{2}\left(q_{1}+k_{1}e^{-\ell q_{0}}\right)^{2}\end{array}\label{joczz1}\end{equation}
while for the right-hand side we obtain
\begin{equation}
\begin{array}{c}
\frac{\partial}{\partial q_{\theta}}\left(q\oplus_{\ell}k\right)_{1}\tilde{N}_{\theta}(q)+e^{-\ell q_{0}}\frac{\partial}{\partial k_{\theta}}\left(q\oplus_{\ell}k\right)_{1}\tilde{N}_{\theta}(k)=\\
=\frac{\partial}{\partial q_{\theta}}\left(q_{1}+k_{1}e^{-\ell q_{0}}\right)\tilde{N}_{\theta}(q)+e^{-\ell q_{0}}\frac{\partial}{\partial k_{\theta}}\left(q_{1}+k_{1}e^{-\ell q_{0}}\right)\tilde{N}_{\theta}(k)=\\
=\left(\tilde{N}_{1}(q)-\ell k_{1}e^{-\ell q_{0}}\tilde{N}_{0}(q)\right)+e^{-2\ell q_{0}}\tilde{N}_{1}(k)=\\
=\frac{1}{2\ell}\left(1-e^{-2\ell\left(q_{0}
+k_{0}\right)}\right)-\frac{\ell}{2}\left(q_{1}+k_{1}e^{-\ell q_{0}}\right)^{2}\end{array}\label{joczz2}\end{equation}
So, as desired, the left-hand side of (\ref{joctwoxi}), which is  (\ref{joczz1}),
agrees with the right-hand side of (\ref{joctwoxi}), which is  (\ref{joczz2}).

This concludes our investigation of the relativistic properties of
the $\kappa$-dS momentum
space.
Evidently
these relativistic properties are technically sound but it does appear that
the condition $k^{[B]} =  {\tilde{\Lambda}}^{\xi e^{-\ell q_{0}} }( k^{[A]})$
may require an additional effort of interpretation.
That condition
essentially implies that the same kind of particle with the same value of momentum $k$
should transform differently depending on whether the particle is freely propagating (no role in any conservation law)
or it enters in an interaction event (so that its momentum $k$ gets composed to other momenta),
with in turn different interaction events producing different transformation laws for $k$.
Dwelling on whether or not such an interpretation leads to \lqt acceptable" physical predictions
goes beyond the scopes of our analysis: the study
of $\kappa$-dS momentum space is a lively research area and we felt
it could be a valuable contribution to its development if we clearly exposed the strengths
and the peculiarities of its relativistic properties. Our findings
for the DSR-relativistic properties of the $\kappa$-dS momentum space
provide a solid technical basis on which future investigations of
the suitability for physics of this momentum space may rely.

While postponing to future studies a more through analysis of the
{}``physical acceptability\textquotedbl{} of the relativistic properties
of the $\kappa$-dS momentum space, we close this section highlighting
a few facts that confirm that at least technically
we do have here a fully relativistic framework.
For this purpose let us start by observing
that the results reported above show that two observers, Alice and Bob,
connected by a pure boost, witnessing energy-momentum conservation
in a particle-physics process, agree on the mass shell to which each
particle belongs and also agree on the form of the law of energy-momentum
conservation (they agree on the form of  $\oplus_\ell$).
Moreover, one should notice that the $\kappa$-dS rules
of transformation between Alice
and Bob are truly relational (no preferred observer), as one easily sees by considering
how the map from Bob to Alice is related to the map from Alice to
Bob. For this take Alice with momenta $q^{[A]}$ and $k^{[A]}$ entering an event through
a composition $q^{[A]} \oplus_{\ell} k^{[A]}$.
And take the map from Alice to Bob to be characterized by rapidity/boost parameter $\xi$,
so that Bob has $q^{[B]} = {\tilde{\Lambda}}^{\xi}( q^{[A]})$
and $k^{[B]} =  {\tilde{\Lambda}}^{\xi e^{-\ell q_{0}^{[A]}} }( k^{[A]})$.
If we now look at the inverse situation, starting with Bob's $q^{[B]} $
and $k^{[B]} $ entering  an event through
a composition $q^{[B]} \oplus_{\ell} k^{[B]}$, then the map from Bob to Alice
is characterized by boost parameter $-\xi$,
and Alice finds $q^{[A]} = {\tilde{\Lambda}}^{-\xi}( q^{[B]})$
and $k^{[A]} =   {\tilde{\Lambda}}^{-\xi e^{-\ell q_{0}^{[B]}}}( k^{[B]})$.
Evidently this does give a consistent picture as
 verified explicitly to leading order in $\xi$ through the following observation
\begin{equation}
\begin{array}{c}
\tilde{\Lambda}^{\xi e^{-\ell q_{0}^{[A]}}}(\tilde{\Lambda}^{-\xi e^{-\ell q_{0}^{[B]}}}(k^{[B]}))\approx\\
\approx\tilde{\Lambda}^{-\xi e^{-\ell q_{0}^{[B]}}}(k^{[B]})+\xi e^{-\ell q_{0}^{[A]}}N^{(\ell)}(\tilde{\Lambda}^{-\xi e^{-\ell q_{0}^{[B]}}}(k^{[B]}))\approx\\
\approx k^{[B]}-\xi e^{-\ell q_{0}^{[B]}}\tilde N(k^{[B]})+\xi e^{-\ell q_{0}^{[A]}}\tilde N(\tilde{\Lambda}^{-\xi e^{-\ell q_{0}^{[B]}}}(k^{[B]}))\approx\\
\approx k^{[B]}-\xi e^{-\ell q_{0}^{[B]}}\tilde N(k^{[B]})+\xi e^{-\ell q_{0}^{[A]}}\tilde N(k^{[B]})=k^{[B]}\end{array}\end{equation}
where we also used the fact that $\xi (q^{[B]} - q^{[A]})=0+O(\xi^2)$.

\section{Relativistic consistency as a geometric property\label{Relativistic consistency as a geometric property}}
In the previous sections we reported some new results that can have direct
applicability in future studies of the $\kappa$-dS, the proper-dS and the weakly-proper-dS momentum spaces.
In this section we want to offer an additional
 contribution to the conceptual characterization
of some of our results,
by considering momentum-space diffeomorphisms.

A significant part of our analysis was focused on  the requirements necessary
for (DSR-)relativistic covariance of theories on momentum space.
It remains unclear whether specifically the formulation of such
theories within the relative-locality framework of Ref.\ttt\cite{principle}
should also be invariant under momentum-space diffeomorphisms, in the sense of
general covariance: this appears at first sight desirable
but several grey areas of understanding remain concerning what happens to these theories
when we change coordinates on momentum space.
Our main objective in this section is to provide evidence of the fact that,
while other properties of momentum-space theories may well depend on
the choice of momentum-space coordinatization,
the property of a momentum space of being DSR-relativistic compatible
is a truly geometric property.
We do this by considering our proper-dS momentum space, with the properties established
within a chosen coordinatization of momentum space\footnote{We note in passing that the coordinatization
of proper-dS momentum space adopted in the previous sections reproduces in leading order in $\ell$ the
results for a DSR-relativistic setup of kinematics reported in Ref.\ttt\cite{dsr1Edsr2}.
In retrospect one can therefore view the results of Ref.\ttt\cite{dsr1Edsr2} as results
at leading order in $\ell$ for the proper-dS momentum space.} in the previous sections
and showing that under the action of a momentum-space diffeomorphism one obtains a different
coordinatization of the proper-dS momentum space which still satisfies the requirements
of DSR-relativistic compatibility.

The results in this section also contribute to the interesting issue,
already explored in Refs.\ttt\cite{GACarXiv11105081,MercatiCarmonaCortes},
that concerns whether the property of theories of being (DSR-)relativistic
can be fully encoded as a momentum-space property.

At the end of this section we also use tools that become available when
considering momentum-space diffeomophisms to establish the
DSR-relativistic compatibility of proper-dS momentum space to all orders in $\ell$
(in the previous sections this was only established at order $\ell^2$).

Because of the nature of the manipulations performed in this section we find convenient
to drop the label $\ell$, leaving the $\ell$-dependence of quantities implicit.
Interested readers can restore $\ell$ by simple use of dimensional analysis.

And, again because of the nature of the manipulations performed in this section,
 we find convenient
to use the notation
$$\tilde\Lambda\left(q\oplus_{\ell} k\right) \equiv \tilde\Lambda\left( {\cal P} \right)\Big|_{{\cal P}=q\oplus_{\ell} k}\ttt.$$
The DSR-deformed boost transformations $\tilde\Lambda$ are functions defined on single momenta,
and we are extending their definition to a momentum obtained by composing two other momenta.\\

\subsection{Diffeomorphisms of proper-dS momentum space}
Let us then consider our proper-dS momentum space, with its symmetry transformation $S$, which
contain in particular the  subset of DSR-deformed boost transformations $\tilde\Lambda$:
\begin{equation}
\begin{cases}
d^{2}\left(S(k),S(q)\right)=d^{2}(k,q)\\
\tilde\Lambda\left(q\oplus_{\Gamma}k\right)=\tilde\Lambda(q)\oplus_{\Gamma}\tilde\Lambda(k)\end{cases}\end{equation}
where we changed notation from $\oplus_\ell$ to  $\oplus_{\Gamma}$ just as a way to stress the role of the
connection $\Gamma$  in the law of composition of momenta. This role of the connection will be at center stage in this section.

Next we consider a diffeomorphism:
\begin{equation}
\begin{cases}
p_{\lambda}\rightarrow p'_{\lambda}=f_{\lambda}(p)\\
V_{\lambda}(p)\rightarrow V'_{\lambda}(p')=J_{\lambda}^{\alpha}V_{\alpha}(p)\\
g^{\mu\nu}(p)\rightarrow g^{'\mu\nu}(p')=\bar{J}_{\alpha}^{\mu}\bar{J}_{\beta}^{\nu}g^{\alpha\beta}(p)\\
\Gamma^{\mu\nu}{}_{\lambda}(p)\rightarrow\Gamma'{}^{\mu\nu}{}_{\lambda}(p')=\bar{J}_{\gamma}^{\nu}\bar{J}_{\rho}^{\mu}J_{\lambda}^{\beta}\Gamma^{\rho\gamma}{}_{\beta}(p)-\bar{J}_{\gamma}^{\nu}\bar{J}_{\rho}^{\mu}\left(\partial^{\rho}J_{\lambda}^{\gamma}\right)\end{cases}\label{eq:diffeo traf rule}\end{equation}
where $V_{\lambda}(p)$ is a generic vector field, $J(p)$ is the
Jacobi matrix $J_{\lambda}^{\alpha}(p)=\partial^{\alpha}f_{\lambda}(p)$
and $\bar{J}$ is the inverse of $J$ .

We are interested in establishing that  in the new coordinates one still
 has a set of symmetries $S'$ and a subset $\tilde\Lambda'$ such that
\begin{equation}
\begin{cases}
d'^{2}\left(S'(k'),S'(q')\right)=d^{2}(k',q')\\
\tilde\Lambda'\left(q'\oplus^{[f(0)]}_{f(\Gamma)}k'\right)=\tilde\Lambda'(q')\oplus^{[f(0)]}_{f(\Gamma)}\tilde\Lambda'(k')\end{cases}\end{equation}
where $d'^{2}$ is the distance function associated to the transformed
metric $g'^{\mu\nu}$ and $\oplus^{[f(0)]}_{f(\Gamma)}$ is the translated composition law that we defined in (\ref{eq:translated comp law}), associated to the subtraction point $f(0)$ and to the transformed connection:
\begin{equation}
f(\Gamma)_{\lambda}^{\mu\nu}=\Gamma'^{\mu\nu}{}_{\lambda}\left(p'\right)
=\bar{J}_{\gamma}^{\nu}\bar{J}_{\rho}^{\mu}J_{\lambda}^{\beta}\Gamma^{\rho\gamma}{}_{\beta}(p)
-\bar{J}_{\gamma}^{\nu}\bar{J}_{\rho}^{\mu}\left(\partial^{\rho}J_{\lambda}^{\gamma}\right)\end{equation}

For what concerns the invariance of the distance function, we can exploit again  the fact
that it is equivalent to considering the invariance of the line element:
\begin{equation}
g^{\mu\nu}(S(p))dS_{\mu}(p)dS_{\nu}(p)=g^{\mu\nu}(p)dp_{\mu}dp_{\nu}\label{eq:onshell invariance}\end{equation}
We then must seek the transformations $S'$ such that
\[
g'^{\mu\nu}(S'(p'))dS'_{\mu}(p')dS'_{\nu}(p')=g'^{\mu\nu}(p')dp'_{\mu}dp'_{\nu}\]
Using the transformation
properties of the metric, of $S$ and of the differential of a vector
we can notice that
\begin{equation}
g'^{\mu\nu}(S'(p'))dS'_{\mu}(p')dS'_{\nu}(p')=g^{\alpha\beta}(S(p))dS{}_{\mu}(p)dS{}_{\nu}(p)\end{equation}
Then, with  (\ref{eq:onshell invariance}) and using again
the transformation properties of the metric we get
\begin{equation}
g^{\alpha\beta}(S(p))dS{}_{\mu}(p)dS{}_{\nu}(p)=g^{\mu\nu}(p)dp_{\mu}dp_{\nu}=g'^{\mu\nu}(p')dp'_{\mu}dp'_{\nu}\end{equation}
which is the result we were seeking.

Our next and final task for establishing that the DSR-relativistic compatibility of proper-dS momentum
space is a diffeomorphism-invariant property concerns the law of composition of momenta.
For this purpose we start from establishing that under a diffeomorphism
the composition law transforms as follows:\begin{equation}
f\left(q\oplus_{\Gamma}k\right)=f(q)\oplus^{[f(0)]}_{f(\Gamma)}f(k)\label{eq:comp diff prop}\end{equation}
We can easily verify this property adopting the formulation of the role of the affine connection
in the composition law which we introduced in Section \ref{sec:geometry}:
\begin{equation}
\begin{cases}
\frac{d\zeta_{\rho}}{dt}\nabla_{\Gamma}^{\rho}\frac{d\gamma_{\lambda}}{ds}=0\\
q\oplus_{\Gamma}k=\gamma(1,1)\end{cases}\label{eq:comp law for diff}\end{equation}
where it should be appreciated that  $\gamma$ is completely determined by the
affine connection $\Gamma$, the points $q$ and $k$ on momentum space and implicitly by the choice of the origin as subtraction point. In particular
we have that
\begin{equation}
\begin{cases}
\gamma(0,1)=q\\
\gamma(1,0)=k\\
\gamma(0,0)=0
\end{cases}\end{equation}
A diffeomorphism $p\rightarrow f(p)$
maps the surface $\gamma$ into a surface $\gamma'$
characterized by
\begin{equation}
\gamma\,'(s,t)=f(\gamma(s,t))
\end{equation}
for which in particular  on has that
\begin{equation}
\begin{cases}
\gamma\,'(0,1)=f(\gamma(0,1))=f(q)\\
$$\gamma\,'(1,0)=f(\gamma(1,0))=f(k)\\
$$\gamma\,'(1,1)=f(\gamma(1,1))=f(q\oplus_{\Gamma}k)\\
\gamma\,'(0,0)=f(0)
\end{cases}\end{equation}

One can observe that, if we change the connection $\Gamma$ and the vectors $\frac{d\zeta_{\rho}}{dt}$
and $\frac{d\gamma\rho}{dt}$
according to  (\ref{eq:diffeo traf rule}),
from (\ref{eq:comp law for diff})
it follows that
\begin{equation}
\frac{d\zeta'_{\rho}}{dt}\nabla_{\Gamma'}^{\rho}\frac{d\gamma'_{\lambda}}{ds}=0
\label{eq:trasformed cov comp}\end{equation}
where the apices refers to the transformed quantities.

Therefore we have that $\gamma'$ is the surface which solves (\ref{eq:trasformed cov comp})
with the transformed connection $\Gamma'$ and whose boundary conditions
are determined by the
points $q'=f(q)$ , $k'=f(k)$ and f(0) as the subtraction point.
So, using the definition (\ref{eq:translated comp law}) of the translated composition law we get that:
\begin{equation}
\gamma\,'(1,1)=f(q)\oplus^{[f(0)]}_{f(\Gamma)}f(k)\end{equation}
which indeed confirms $f\left(q\oplus_{\Gamma}k\right)=f(q)\oplus^{[f(0)]}_{f(\Gamma)}f(k)$, as announced
in (\ref{eq:comp diff prop}).

This result provides the key ingredient for us to show that, as announced
\begin{equation}
\tilde\Lambda'\left(q'\oplus^{[f(0)]}_{f(\Gamma)}k'\right)=\tilde\Lambda'(q')\oplus^{[f(0)]}_{f(\Gamma)}\tilde\Lambda'(k')
\label{jocjuice}
\end{equation}
For this purpose we observe that
\begin{equation}
\tilde\Lambda'(p')=f(\tilde\Lambda(p))\label{eq: diff lorentz transform}\end{equation}
from which, using (\ref{eq:comp diff prop}) and (\ref{eq: diff lorentz transform}),
we obtain
\begin{equation}
\tilde\Lambda'\left(q'\oplus^{[f(0)]}_{f(\Gamma)}k'\right)
=\tilde\Lambda'(f\left(q\oplus_{\Gamma}k\right))
=f\left(\tilde\Lambda\left(q\oplus_{\Gamma}k\right)\right)\end{equation}
We then use the  covariance of the composition law that we assume to
hold for the untransformed coordinates
\begin{equation}
f\left(\tilde\Lambda\left(q\oplus_{\Gamma}k\right)\right)=f\left(\tilde\Lambda(q)\oplus_{\Gamma}\tilde\Lambda(k)\right)\end{equation}
again in combination with
 (\ref{eq:comp diff prop}) and (\ref{eq: diff lorentz transform}), to establish that
\begin{equation}
f\left(\tilde\Lambda(q)\oplus_{\Gamma}\tilde\Lambda(k)\right)
=f\left(\tilde\Lambda(q)\right)\oplus^{[f(0)]}_{f(\Gamma)}f\left(\tilde\Lambda(k)\right)=\tilde\Lambda'(q')\oplus^{[f(0)]}_{f(\Gamma)}\tilde\Lambda'(k')\end{equation}
The last three equations confirm the validity of (\ref{jocjuice}),
and this completes our proof of the fact that the DSR-relativistic compatibility of proper-dS momentum
space is a diffeomorphism-invariant property.

\subsection{All-order DSR-relativistic compatibility of proper-dS momentum space\label{sec:general standard rel cond}}
Some of these techniques based on diffeomorphism transformations
also allow us to establish, as announced, the DSR-relativistic compatibility of
proper-dS momentum space to all orders in $\ell$.
In this section we are keeping the $\ell$ dependence implicit, but the validity 
of results to all order
in $\ell$ will of course be evident from the fact that we establish exact results.

Let us start building this proof by first using some results of the previous subsection
for introducing some requirements that ensure the DSR-relativistic compatibility
 of the composition law. The first requirement is that
 the origin of momentum space is invariant under boosts ($f(0)=0$), as in the proper-dS case. Using (\ref{eq:comp diff prop}) this ensures that
\begin{equation}
\tilde\Lambda\left(q\oplus_{\Gamma}k\right)=\tilde\Lambda\left(q\right)\oplus_{\Gamma'}\tilde\Lambda\left(k\right)\label{eq:actionoflambda}\end{equation}
where $\Gamma'$ is 
\begin{equation}
\Gamma'^{\mu\rho}{}_{\alpha}\left(p'\right)
\equiv \bar{J}_{\gamma}^{\rho}\bar{J}_{\nu}^{\mu}J_{\alpha}^{\beta}\Gamma^{\nu\gamma}{}_{\beta}(p)
-\bar{J}_{\gamma}^{\rho}\bar{J}_{\nu}^{\mu}\left(\partial^{\nu}J_{\alpha}^{\gamma}\right)\end{equation}
 $J$ is the Jacobian $J_{\alpha}^{\mu}=\partial^{\mu}\tilde\Lambda_{\alpha}$
and $\bar{J}$ is the inverse of $J$. The other requirement, also evidently satisfied in the proper-dS case, 
is the invariance of the connection under boosts ($\Gamma'=\Gamma$). When both requirements are satisfied
the covariance of the composition law is indeed guaranteed, since combining invariance of the connection
with (\ref{eq:actionoflambda}) one has that
\begin{equation}
\Gamma'=\Gamma \rightarrow \tilde\Lambda\left(q\oplus_{\Gamma}k\right)=\tilde\Lambda\left(q\right)\oplus_{\Gamma}\tilde\Lambda\left(k\right)\label{eq:invofcon covofcomp}\end{equation}

Having established this powerful point about the DSR-relativistic compatibility of composition laws, let us now
consider the case of  a suitable momentum space metric $g$, such that
 a DSR-deformed boost $\tilde\Lambda$ is a diffeomorphism
on momentum space with the property
\begin{equation}\label{eq:on-shell preserving boosts}
d^{2}\left(\tilde\Lambda(p),\tilde\Lambda(q)\right)=d^{2}\left(p,q\right) \, .\end{equation}
which is equivalent to the condition of invariance of the metric:
\begin{equation}
g^{\mu\nu}\left(p'\right)=g'^{\mu\nu}(p')
=\bar{J}_{\alpha}^{\mu}\bar{J}_{\beta}^{\nu}g^{\alpha\beta}(p)\label{eq:metric invriance}\end{equation}
If then, as in the proper-dS case, one adopts as  composition law
 the one produced, according to our novel geometric interpretation, by
 the Levi-Civita connection of the metric $g$
\begin{equation}
A{}^{\mu\rho}{}_{\alpha}\left(p\right)
=\frac{1}{2}\bar{g}_{\alpha\beta}(p)\left(g^{\beta\mu,\rho}(p)+g^{\beta\rho,\mu}(p)-g^{\mu\rho,\beta}(p)\right)\label{eq:Levi Civita definition2}\end{equation} 
the overall DSR-relativistic compatibility is guaranteed. In fact, it is easy to see that the Levi-Civita connection is always invariant under the  deformed boosts $\tilde\Lambda$,
\begin{equation}
A'^{\mu\rho}{}_{\alpha}\left(p'\right)
=\frac{1}{2}\bar{g}'_{\alpha\beta}(p')\left(g'^{\beta\mu,\rho}(p')+g'^{\beta\rho,\mu}(p')-g'^{\mu\rho,\beta}(p')\right)
=\frac{1}{2}\bar{g}_{\alpha\beta}(p')\left(g^{\beta\mu,\rho}(p')+g^{\beta\rho,\mu}(p')-g^{\mu\rho,\beta}(p')\right)
=A^{\mu\rho}{}_{\alpha}(p') \, ,
\end{equation}
which in turn, using (\ref{eq:invofcon covofcomp}), leads us to our sought result:
\begin{equation}
\tilde\Lambda\left(q\right)\oplus_{A'}\tilde\Lambda\left(k\right)
=\tilde\Lambda\left(q\right)\oplus_{A}\tilde\Lambda\left(k\right)
\end{equation}

This concludes our proof of exact (DSR-)relativistic invariance. And it should be noticed that
in this subsection we did not make use of any result based on the specific form of
the dS metric and/or its Levi-Civita connection.
So evidently our proof of (DSR-)relativistic invariance applies to a wider class of momentum spaces:
when adopting the novel interpretation of the composition law which we here introduced 
in (\ref{eq: comp from conn})-(\ref{eq: comp is the extr}) one gets a DSR-relativistic-compatible setup
whenever the metric is DSR-compatible (in the sense of (\ref{eq:on-shell preserving boosts}))
and the composition law is formulated using
(in the sense of the novel geometric interpretation we here introduced
through Eqs. (\ref{eq: comp from conn})-(\ref{eq: comp is the extr}))
its associated Levi-Civita connection.
Certain other  choices of momentum-space affine connection are also
DSR-relativistic compatible with such a momentum-space metric (see, {\it e.g.}, the case of $\kappa$-dS momentum
space), but the specific possibility of pairing the metric with its Levi-Civita connection
ensures the overall DSR-relativistic compatibility.
We feel that this should be viewed as an aspect of compellingness of our geometric interpretation
centered on (\ref{eq: comp from conn})-(\ref{eq: comp is the extr}).

\section{Outlook}
In closing
we offer a few remarks on what we see as the most significant potential
implications of the
results we here reported.

The most direct implications are for the study of the recently-proposed
relative-locality framework.
The momentum space which has been so far most studied within this framework
is the $\kappa$-dS momentum
space. And we here provided several additional tools for the
investigation of this possibility,
particularly through our
analysis of relativistic properties,
which went in greater depth than any such previous analysis of the
$\kappa$-dS momentum
space.

The appeal of the $\kappa$-dS momentum space resides (among other things)
in the associativity of the composition law.
But the affine connections that provide associative composition laws
are a small subset of those one could in principle consider for
the relative-locality framework, so in that respect the $\kappa$-dS momentum space
is not representative of the possible structure of theories on curved momentum spaces.
Moreover, the $\kappa$-dS momentum space has a noncommutative law of composition of momenta,
which is not a general property of curved momentum spaces and is also not necessarily
a ``desirable" property (the associated interpretational challenges can be handled~\cite{principle},
but indeed they are not necessarily desirable).

The proper-dS and the weakly-proper-dS momentum spaces which we here introduced are usefully
complementary to the $\kappa$-dS momentum space, at least in the sense that they have non-associative
but commutative laws of composition of momenta. We were here primarily focusing on the structure and
relativistic properties of these momentum spaces. Future works should establish what are the observable
differences between cases with associative but noncommutative
composition laws and cases with non-associative but commutative
composition laws.

These new cases with non-associative but commutative
composition laws also deserve interest for the novelty of the formalization
of their DSR-relativistic symmetries.
It is rather well understood that in cases like $\kappa$-dS 
the momentum space is a group manifold, and the formalization
of the
DSR-relativistic symmetries can be based on the mathematics of some corresponding Hopf algebras,
then leading inevitably to
non-associative but commutative
composition laws.
It would be important to understand what sort of general mathematical structures should replace
Hopf algebras in cases where the momentum space is DSR-relativistic
but the composition law is commutative
but
non-associative.

Perhaps the most fascinating issue we here contemplated
concerns the general possibility of associating firmly a momentum-space geometry
to a given setup for Planck-scale-deformed relativistic kinematics.
We here established
that the possible choices of laws of composition of momenta are more numerous than
the possible choices of affine connection on a momentum space.
This inevitably raises the issue of finding a most natural way for introducing
a link between affine connection and form of the composition law.
As a first contribution toward tackling this issue
we here offered several observations concerning
the differences between
 the standard geometric-interpretation prescription, here
coded in Eq.(\ref{eq: Connection Introduction}), and the novel geometric interpretation we introduced
through Eqs.  (\ref{eq: comp from conn})-(\ref{eq: comp is the extr}).

\appendix

\section{Decomposition of the Connection\label{sec:Decomposition of the Connection}}

A well known result of differential geometry states that, given a connection
$\Gamma$ and a metric $g$ , the connection can be decomposed as:

\begin{equation}
\Gamma^{\lambda\mu}{}_{\nu}=A^{\lambda\mu}{}_{\nu}+K^{\lambda\mu}{}_{\nu}+V^{\lambda\mu}{}_{\nu}\end{equation}

$A^{\lambda\mu}\,_{\nu}$ is the Levi-Civita connection, defined by the requirement that the covariant derivative associated to it  vanishes when applied to the metric tensor:

\begin{equation}
\begin{cases}
\nabla_{(A)}^{\lambda}g^{\mu\nu}=\partial^{\lambda}g^{\mu\nu}-A^{\lambda\mu}{}_{\theta}g^{\theta\nu}-A^{\lambda\nu}{}_{\theta}g^{\theta\mu}=0\\
A^{\lambda\mu}{}_{\nu}=A^{\mu\lambda}{}_{\nu}\end{cases}\label{eq:Levi Civita definition}\end{equation}

This system is solved by:

\begin{equation}
A^{\lambda\mu}{}_{\nu}=\frac{1}{2}{g}_{\nu\theta}\left(\partial^{\mu}g^{\theta\lambda}+\partial^{\lambda}g^{\theta\mu}-\partial^{\theta}g^{\lambda\mu}\right)\end{equation}

$K^{\lambda\mu}\,_{\nu}$ is called the contortion tensor and is defined by:

\begin{equation}
K^{\lambda\mu}{}_{\theta}g^{\theta\nu}+K^{\lambda\nu}{}_{\theta}g^{\theta\mu}=0\end{equation}

while $V^{\lambda\mu}\,_{\nu}$ is the cononmetricity, defined by:

\begin{equation}
\nabla_{(\Gamma)}^{\lambda}g^{\mu\nu}=-V^{\lambda\mu}{}_{\theta}g^{\theta\nu}-V^{\lambda\nu}{}_{\theta}g^{\theta\mu}\end{equation}
where $\nabla_{(\Gamma)}$ is the covariant derivative associated to the connection $\Gamma$.

In order to express  $K$ and $V$  in terms of the metric $g$ and the connection $\Gamma$
it is convenient to make use of the torsion tensor and the nonmetricity tensor, defined respectively as:

\[
\begin{cases}
T^{\lambda\mu}{}_{\nu}=2\Gamma^{[\lambda\mu]}{}_{\nu}\\
Q^{\lambda\mu\nu}=\nabla^{\nu}g^{\lambda\mu}\end{cases}\]

Observing that:

\begin{equation}
\left(\partial^{\lambda}g^{\mu\nu}-\partial^{\nu}g^{\lambda\mu}-\partial^{\mu}g^{\nu\lambda}\right)-2\Gamma^{[\lambda\mu]}{}_{\theta}g^{\theta\nu}-2\Gamma^{[\lambda\nu]}{}_{\theta}g^{\theta\mu}+2\Gamma^{(\mu\nu)}{}_{\theta}g^{\theta\lambda}=Q^{\mu\nu\lambda}-Q^{\lambda\mu\nu}-Q^{\nu\lambda\mu}\end{equation}
and using the definition of the torsion tensor, one finds:

\begin{equation}
\Gamma^{\mu\nu\lambda}=A^{\mu\nu\lambda}+\frac{1}{2}\left(T^{\mu\nu\lambda}+T^{\lambda\mu\nu}-T^{\nu\lambda\mu}\right)+\frac{1}{2}\left(Q^{\mu\nu\lambda}-Q^{\lambda\mu\nu}-Q^{\nu\lambda\mu}\right)\end{equation}

Finally, lowering one index:

\begin{equation}
\Gamma^{\mu\nu}{}_{\lambda}=A^{\mu\nu}{}_{\lambda}+\frac{1}{2}\left(T^{\mu\nu}{}_{\lambda}+T_{\lambda}{}^{\mu\nu}-T^{\nu}{}_{\lambda}{}^{\mu}\right)+\frac{1}{2}\left(Q^{\mu\nu}{}_{\lambda}-Q{}_{\lambda}{}^{\mu\nu}-Q^{\nu}{}_{\lambda}{}^{\mu}\right)\end{equation}

The second and
the third term in the above expression satisfy  the defining properties
of, respectively, $K$ and $V$. Therefore we  define:

\begin{equation}
\begin{cases}
K^{\mu\nu}{}_{\lambda}=\frac{1}{2}\left(T^{\mu\nu}{}_{\lambda}+T_{\lambda}{}^{\mu\nu}-T^{\nu}{}_{\lambda}{}^{\mu}\right)\\
V^{\mu\nu}{}_{\lambda}=\frac{1}{2}\left(Q^{\mu\nu}{}_{\lambda}-Q{}_{\lambda}{}^{\mu\nu}-Q^{\nu}{}_{\lambda}{}^{\mu}\right)\end{cases}\end{equation}

From these definitions one easily verifies that:

\begin{equation}
\begin{cases}
K^{\mu(\nu\lambda)}=\frac{1}{2}\left(T^{\mu(\nu\lambda)}+T^{(\lambda\mu\nu)}\right)=0\\
V^{[\mu\nu]\lambda}=0=Q^{[\mu\nu]\lambda}\end{cases}\end{equation}

Therefore, the symmetries of the contortion do not involve the differential
index, while the symmetries of the cononmetricity do, as is the case for the Levi-Civita connection.

\section{Compatibility conditions for the invariance of the distance function}\label{sec:cov of distance app}

We want to find the symmetries of the distance function

\begin{equation}
d^{2}_{\ell}\left(k,q\right)=\int dt\sqrt{g^{\mu\nu}(\gamma(t))\dot{\gamma}_{\mu}(t)\dot{\gamma}_{\nu}(t)}\label{eq:distance function app}\end{equation}

where $\gamma(t)$ is the metric geodesic connecting  the points $q$ and $k$.

So we look for the set of infinitesimal transformations on momentum space
$p_{\mu}\rightarrow{\cal S}_{\mu}^{(\ell)}(p)=p_{\mu}+{\cal T}_{\mu}^{(\ell)}(p)$
 which act on the two points $q$ and $k$ preserving the geodesic distance between them:

\begin{equation}
d_{\ell}^{2}(k+{\cal T}^{(\ell)}(k),q+{\cal T}^{(\ell)}(q))=d_{\ell}^{2}(k,q)\label{eq: on-shell inf invariance}\end{equation}

Thanks to the monotone behavior of the square root we can equivalently ask that the following quantity is left unchanged by the transformations:
\begin{equation}
\tilde{d}_{\ell}\left(k,q\right)=\int dt\left(g^{\mu\nu}(\gamma(t))\dot{\gamma}_{\mu}(t)\dot{\gamma}_{\nu}(t)\right)\label{eq:squared distance}\end{equation}

Defining $L(\gamma,\dot{\gamma})\equiv g^{\mu\nu}(\gamma)\dot{\gamma}_{\mu}\dot{\gamma}_{\nu}$,
 the variation of (\ref{eq:squared distance}) takes the form:

\begin{equation}
\begin{array}{c}
\delta\int dtL(\gamma(t),\dot{\gamma}(t))=\int\left(\frac{\partial L}{\partial\gamma}-\frac{d}{dt}\left(\frac{\partial L}{\partial\dot{\gamma}}\right)\right)\delta\gamma dt+\left(\frac{\partial L}{\partial\dot{\gamma}}\delta\gamma\right)\bigg |_{t=0}^{t=1}\end{array}\label{eq:variation of distance}\end{equation}

where $\delta\gamma(t)$ is such that $\gamma(t)+\delta\gamma(t)$ is the geodesic connecting $q+{\cal T}(q)$ and $k+{\cal T}(k)$, which must satisfy, in particular, $\delta\gamma(0)={\cal T}(q)$
and $\delta\gamma(1)={\cal T}(k)$.

The argument of the integral on the right-hand side of ($\ref{eq:variation of distance}$) is  nothing but the geodesic equation which is solved, by definition, by the curve $\gamma$. So this contribution to the variation of the distance function  vanishes and we  only need to ask for the boundary term to vanish:

\begin{equation}
\left(\frac{\partial L}{\partial\dot{\gamma}}\delta\gamma\right)\bigg|_{t=0}^{t=1}=0\label{eq:on shell equation}\end{equation}

In order to solve (\ref{eq:on shell equation}) the first step is to note that, since we have to solve 
(\ref{eq:on shell equation}) for any couple of points $q$ and $k$,
 we can equivalently solve:

\begin{equation}
\frac{d}{dt}\left(\frac{\partial L}{\partial\dot{\gamma}}\delta\gamma\right)=0\end{equation}

This implies that the Killing equation must be satisfied by $\delta\gamma$:
\begin{equation}
\begin{array}{c}
\frac{d}{dt}\left(\frac{\partial L}{\partial\dot{\gamma}}\delta\gamma\right)=\left(\partial^{\theta}g^{\alpha\nu}\delta\gamma_{\alpha}-g^{\alpha\eta}A_{\eta}^{\theta\nu}\delta\gamma_{\alpha}+g^{\alpha\nu}\partial^{\theta}\delta\gamma_{\alpha}\right)\dot{\gamma}_{\theta}\dot{\gamma}_{\nu}\\
\Rightarrow 0=\left(\partial^{(\theta}g^{\alpha\nu)}\delta\gamma_{\alpha}-g^{\alpha\eta}A_{\eta}^{\theta\nu}\delta\gamma_{\alpha}+g^{\alpha(\nu}\partial^{\theta)}\delta\gamma_{\alpha}\right)=\left(\partial^{(\theta}\delta\gamma^{\nu)}-A_{\eta}^{\theta\nu}\delta\gamma^{\eta}\right)
\end{array}
\label{eq:computations}\end{equation}

that is:

\begin{equation}
\nabla_{A}^{(\theta}\delta\gamma^{\nu)}=0\label{eq:delta gamma kills it}\end{equation}

By identifying $\delta\gamma$ with ${\cal T}^{(\ell)}$ \footnote{It is possible to show that this identification is  a necessary condition in order for (\ref{eq:variation of distance}) to hold.} we  get the condition on the infinitesimal transformations $\cal T$ ensuring the invariance of the distance function:

\begin{equation}
\nabla_{A}^{(\theta}{\cal {\cal T}^{(\ell)\nu)}}=0\label{eq:Killing equation for T app}\end{equation}

We now turn to the task of finding the explicit form of the symmetries
${\cal T}^{(\ell)}$ which preserve the distance function associated
to the de Sitter metric.

Using
 the de Sitter Levi-Civita connection

\begin{equation}
A^{\lambda\mu}{}_{\nu}(p)=2\ell\delta_{0}^{(\lambda}\delta_{1}^{\mu)}\delta_{\nu}^{1}+\ell\delta_{1}^{\lambda}\delta_{1}^{\mu}\delta_{\nu}^{0}e^{\ell p_{0}}\end{equation}
  the condition  (\ref{eq:Killing equation for T app}) becomes

\begin{equation}
\begin{cases}
\partial^{0}{\cal T}^{(\ell)0}=0\\
\frac{1}{2}\left(\partial^{0}{\cal T}^{(\ell)1}+\partial^{1}{\cal T}^{(\ell)0}\right)-\ell{\cal T}^{(\ell)1}=0\\
\partial^{1}{\cal T}^{(\ell)1}-\ell e^{\ell p_{0}}{\cal T}^{(\ell)0}=0\end{cases}\end{equation}

and has the solution:

\begin{equation}
\begin{cases}
{\cal T}_{0}^{(\ell)}(p)=\xi p_{1}+\gamma\\
{\cal T}_{1}^{(\ell)}(p)=-\frac{\xi}{2\ell}e^{-2\ell p_{0}}-\frac{\ell}{2}\xi p_{1}^{2}-\ell\gamma p_{1}+\beta\end{cases}\end{equation}

where $\xi$ , $\beta$ and $\gamma$ are constants of integration.

By requiring that in the limit $\ell\rightarrow0$ we recover the
Poincar\'e group of trasformation we redefine: $\beta\rightarrow\frac{\xi}{2\ell}+\beta$,
obtaining:

\begin{equation}
\begin{cases}
{\cal T}_{0}^{(\ell)}(p)=\xi p_{1}+\gamma\\
{\cal T}_{1}^{(\ell)}(p)=\frac{\xi}{2\ell}\left(1-e^{-2\ell p_{0}}\right)-\xi\frac{\ell}{2} p_{1}^{2}-\gamma \ell p_{1}+\beta\end{cases}\end{equation}

Therefore we get  three killing vectors associated to
the three constants of integration. In particular we recognize a deformation
of the translational sector in the algebra associated to $\beta$ and
$\gamma$ , while the deformation of the Lorentz sector ( ${\cal N^{(\ell)}}\equiv{\cal T}^{(\ell)}|_{\text{Lorentz}}$ that solves
${\cal N}^{(\ell)}(0)=0$) is:

\begin{equation}
\begin{cases}
{\cal N}_{0}^{(\ell)}(p)=\xi p_{1}\\
{\cal N}_{1}^{(\ell)}(p)=\frac{\xi}{2\ell}\left(1-e^{-2\ell p_{0}}\right)-\xi \frac{\ell}{2} p_{1}^{2}\end{cases}\end{equation}

In particular, to the second order in $\ell$ one finds:

\begin{equation}
\begin{cases}
{\cal N}_{0}^{(\ell)}=\xi p_{1}\\
{\cal N}_{1}^{(\ell)}(p)=\xi p_{0}-\xi \ell\left( p_{0}^{2}+\frac{1}{2}p_{1}^{2}\right)+\frac{2}{3}\xi\ell^{2}p_{0}^{3}\end{cases}\end{equation}

\section{Second Order Composition Law\label{sec:Second Order Composition Law}}

We want to develop the form of the composition law associated to some
given connection $\Gamma$ at second order in the Planck length $\ell$.

Our starting point is the definition of the composition law associated
to a given connection, which we note down here again: \begin{equation}
\begin{cases}
\frac{d}{dt}\frac{d}{ds}\gamma_{\lambda}(s,t)+\Gamma^{\alpha\beta}\,_{\lambda}(\zeta^{(s)}(t))\frac{d\zeta^{(s)}_{\alpha}(t)}{dt}\frac{d\gamma_{\beta}(s,t)}{ds}=0\\
\gamma_{\lambda}(s,0)=\gamma_{\lambda}^{(k)}(s)\\
\gamma_{\lambda}(0,t)=\gamma_{\lambda}^{(q)}(t)\\
\gamma_{\lambda}(1,1)=\left(q\oplus_{\ell}k\right)_{\lambda}\end{cases}\label{eq:comp law app}\end{equation}
 where $\zeta$ is a geodesic defined by: \begin{equation}
\begin{cases}
\ddot{\zeta}^{(s)}_{\lambda}(t)+\Gamma^{\alpha\beta}\,_{\lambda}\dot{\zeta}^{(s)}_{\alpha}(t)\dot{\zeta}^{(s)}_{\beta}(t)=0\\
\zeta^{(s)}_{\lambda}(0)=\gamma_{\lambda}(s,0)=\gamma^{(k)}_{\lambda}(s)\\
\zeta^{(s)}_{\lambda}(1)=\gamma_{\lambda}(s,1)\end{cases}\label{eq:xi definition app}\end{equation}

The geodesics $\zeta^{(s)}$, $\gamma^{(k)}$ and $\gamma^{(q)}$
are all connection geodesics which reduce to metric geodesics when
the connection $\Gamma$ has only the Levi-Civita contribution $A$.

The second order solution is found by  expanding the connection as follows:
\begin{equation}
\Gamma^{\lambda\mu}{}_{\nu}(\zeta^{(s)}(t))=\Gamma^{\lambda\mu}{}_{\nu}(0)+\partial^{\theta}\Gamma^{\lambda\mu}{}_{\nu}|_{\zeta=0}\zeta_{\theta}^{(s)}(t)\end{equation}
 where the $\zeta^{(s)}(t)$ on the right hand side has to be taken at the zeroth order. In the following, in order to unburden the notation we
will use :

\begin{equation}
\begin{cases}
\Gamma^{\lambda\mu}{}_{\nu}(0)\equiv \tilde \Gamma^{\lambda\mu}{}_{\nu}\\
\partial^{\eta}\Gamma^{\lambda\mu}{}_{\nu}|_{\zeta=0}\equiv\partial^{\eta} \tilde \Gamma^{\lambda\mu}{}_{\nu}\end{cases}\end{equation}
Note that this notation is slightly different from the one used in the main text of the paper in section \ref{sec:Differences}.

Let us now compute the composition law defined by the system of equations
(\ref{eq:comp law app}) and (\ref{eq:xi definition app}).

At the zeroth order
\begin{equation}
\begin{cases}
\frac{d}{dt}\frac{d}{ds}\gamma_{\lambda}(s,t)=0\\
\ddot{\zeta}^{(s)}_{\lambda}(t)=0\end{cases}\end{equation}
which, taking into account the boundary conditions, have solutions:
\begin{equation}
\begin{cases}
\gamma_{\lambda}(s,t)=q_{\lambda}t+k_{\lambda}s\\
\zeta_{\lambda}^{(s)}(t)=q_{\lambda}t+k_{\lambda}s\end{cases}
\end{equation}
So up to the zeroth order the composition law is:
\begin{equation}
\left(q\oplus_{\ell}k\right)_{\lambda}=q_{\lambda}+k_{\lambda}\end{equation}

At the first order in $\ell$ the differential equation in (\ref{eq:comp law app}) reads:
\begin{equation}
\frac{d}{dt}\frac{d}{ds}\gamma_{\lambda}(s,t)
+\tilde\Gamma^{\alpha\beta}\,_{\lambda}\frac{d\zeta_{\alpha}^{(s)}(t)}{dt}\frac{d\gamma_{\beta}(s,t)}{ds}=0\end{equation}
where $\frac{d\zeta_{\alpha}^{(s)}(t)}{dt}$ and $\frac{d\gamma_{\beta}(s,t)}{ds}$ have to be taken at zeroth order, so that the equation takes the form:
\begin{equation}
\frac{d}{dt}\frac{d}{ds}\gamma_{\lambda}(s,t)+\tilde\Gamma^{\alpha\beta}\,_{\lambda}q_{\alpha}k_{\beta}=0\end{equation}

Solving for $\gamma_{\lambda}(s,t)$ and taking into account the boundary conditions we find that up to first order:
\begin{equation}
\gamma_{\lambda}(s,t)=\gamma_{\lambda}^{(q)}(t)+\gamma_{\lambda}^{(k)}(s)-\tilde\Gamma^{\alpha\beta}\,_{\lambda}q_{\alpha}k_{\beta}st\end{equation}
so that the first-order  composition law takes the form:
\begin{equation}
\left(q\oplus_{\ell}k\right)_{\lambda}=q_{\lambda}+k_{\lambda}-\tilde\Gamma^{\alpha\beta}\,_{\lambda}q_{\alpha}k_{\beta}\end{equation}

For what concerns the first-order description of the geodesic $\zeta^{(s)}(t)$ we have that it must satisfy the following differential equation:
\begin{equation}
\ddot{\zeta}^{(s)}_{\lambda}+\tilde\Gamma^{\alpha\beta}\,_{\lambda}\dot{\zeta}_{\alpha}^{(s)}\dot{\zeta}_{\beta}^{(s)}=0\end{equation}

which, together with the boundary conditions, has solution:

\begin{equation}
\zeta_{\lambda}^{(s)}(t)=\gamma_{\lambda}^{(k)}(s)+\gamma_{\lambda}^{(q)}(t)-\tilde\Gamma^{\alpha\beta}\,_{\lambda}q_{\alpha}k_{\beta}st\end{equation}

The second-order contribution to the differential equation in (\ref{eq:comp law app}) is:

\begin{eqnarray}
&&\left(\frac{d}{dt}\frac{d}{ds}\gamma_{\lambda}(s,t)\right)^{(2)}+\tilde\Gamma^{\alpha\beta}\,_{\lambda}\left(\frac{d\zeta_{\alpha}^{(s)}(t)}{dt}\right)^{(0)}\left(\frac{d\gamma_{\beta}(s,t)}{ds}\right)^{(1)}+\tilde\Gamma^{\alpha\beta}\,_{\lambda}\left(\frac{d\zeta_{\alpha}^{(s)}(t)}{dt}\right)^{(1)}\left(\frac{d\gamma_{\beta}(s,t)}{ds}\right)^{(0)}\nonumber \\
&&-\tilde\Gamma^{\alpha\beta}\,_{\lambda}\left(\frac{d\zeta_{\alpha}^{(s)}(t)}{dt}\right)^{(0)}\left(\frac{d\gamma_{\beta}(s,t)}{ds}\right)^{(0)}+\partial^{\theta}\tilde\Gamma^{\alpha\beta}\,_{\lambda}\zeta_{\theta}\left(\frac{d\zeta_{\alpha}^{(s)}(t)}{dt}\right)^{(0)}\left(\frac{d\gamma_{\beta}(s,t)}{ds}\right)^{(0)}=0\end{eqnarray}
where the superscripts refer to the $\ell$ order to be considered for each term.

Substituting the zeroth order results and integrating one finds:

\begin{equation}
\begin{array}{lcl}
\gamma_{\lambda}(s,t)&=&B_{\lambda}(t)+\int ds'A_{\lambda}(s)-\tilde\Gamma^{\alpha\beta}\,_{\lambda}q_{\alpha}\int dt'\int ds'\left(\frac{d\gamma_{\beta}}{ds}\right)^{(1)}-\tilde\Gamma^{\alpha\beta}\,_{\lambda}\int ds'\int dt'\left(\frac{d\zeta_{\alpha}}{dt}\right)^{(1)}k_{\beta}+\\
&&+\tilde\Gamma^{\alpha\beta}\,_{\lambda}q_{\alpha}k_{\beta}st-\frac{1}{2}\partial^{\theta}\tilde\Gamma^{\alpha\beta}{}_{\lambda}\left(q_{\theta}st^{2}+k_{\theta}s^{2}t\right)q_{\alpha}k_{\beta}\end{array}\end{equation}
Then observing that
\begin{equation}
\begin{cases}
\int ds'\left(\frac{d\gamma_{\beta}}{ds}\right)^{(1)}=\gamma_{\beta}^{(k)}(s)-\tilde\Gamma^{\gamma\delta}\,_{\beta}q_{\gamma}k_{\delta}st\\
\int dt'\left(\frac{d\zeta_{\alpha}^{s}}{dt}\right)^{(1)}=\gamma_{\alpha}^{(q)}(t)-\tilde\Gamma^{\gamma\delta}\,_{\alpha}q_{\gamma}k_{\delta}st\end{cases}\end{equation}
and using  the boundary conditions one obtains that up to second order

\begin{equation}
\begin{array}{lcl}
\gamma_{\lambda}(s,t)&=&\gamma_{\lambda}^{(q)}(t)+\gamma_{\lambda}^{(k)}(s)-\tilde\Gamma^{\alpha\beta}\,_{\lambda}q_{\alpha}\left(\left(\gamma_{\beta}^{(k)}(s)\right)^{(1)}t-\frac{1}{2}\tilde \Gamma^{\gamma\delta}\,_{\beta}\,q_{\gamma}k_{\delta}s\,t^{2}\right)+\\
&&-\tilde\Gamma^{\alpha\beta}\,_{\lambda}\left(\left(\gamma_{\alpha}^{(q)}(t)\right)^{(1)}s-\frac{1}{2}\tilde\Gamma^{\gamma\delta}\,_{\alpha}q_{\gamma}k_{\delta}s^{2}t\right)k_{\beta}+\tilde\Gamma^{\alpha\beta}\,_{\lambda}q_{\alpha}k_{\beta}st-\frac{1}{2}\partial^{\theta}\tilde\Gamma^{\alpha\beta}{}_{\lambda}\left(q_{\theta}st^{2}+k_{\theta}s^{2}t\right)q_{\alpha}k_{\beta}\end{array}\end{equation}
So up to second order the composition law $q\oplus_{\ell}k$ associated to a connection $\Gamma$
takes the form:
\begin{equation}
\left(q\oplus_{\ell}k\right)_{\lambda}=q_{\lambda}+k_{\lambda}-\tilde\Gamma^{\alpha\beta}\,_{\lambda}q_{\alpha}k_{\beta}+\frac{1}{2}\tilde\Gamma^{\alpha\beta}\,_{\lambda}\tilde\Gamma^{\gamma\delta}\,_{\beta}q_{\alpha}q_{\gamma}k_{\delta}+\frac{1}{2}\tilde\Gamma^{\alpha\beta}\,_{\lambda}\tilde\Gamma^{\gamma\delta}\,_{\alpha}q_{\gamma}k_{\beta}k_{\delta}-\frac{1}{2}\partial^{\theta}\tilde\Gamma^{\alpha\beta}\,_{\lambda}\left(q_{\theta}+k_{\theta}\right)q_{\alpha}k_{\beta}\end{equation}

If we use, for example, the Levi-Civita connection associated
to the de Sitter metric
\begin{equation}
\Gamma^{\lambda\mu}{}_{\nu}(p)=\ell\left(\delta_{0}^{\lambda}\delta_{1}^{\mu}+\delta_{0}^{\mu}\delta_{1}^{\lambda}\right)\delta_{\nu}^{1}+\ell\delta_{1}^{\lambda}\delta_{1}^{\mu}\delta_{\nu}^{0}e^{\ell p_{0}} \,\end{equation}
 we get the second order
composition law for {}``proper de Sitter'' :
\begin{equation}
\begin{cases}
\left(q\oplus_{\ell}k\right)_{0}=q_{0}+k_{0}-\ell q_{1}k_{1}+\frac{\ell^{2}}{2}\left[-q_{1}k_{1}\left(q_{0}+k_{0}\right)+q_{0}k_{1}^{2}+q_{1}^{2}k_{0}\right]\\
\left(q\oplus_{\ell}k\right)_{1}=q_{1}+k_{1}-\ell\left(q_{0}k_{1}+q_{1}k_{0}\right)+\frac{\ell^{2}}{2}\left[\left(q_{0}k_{1}+q_{1}k_{0}\right)\left(q_{0}+k_{0}\right)+q_{1}k_{1}^{2}+q_{1}^{2}k_{1}\right]\end{cases}\label{eq:second propds comp law app}\end{equation}

\section{Second Order Translated Composition Law\label{sec:P Comp Law}}

The translated composition law associated
to the subtraction point $p$ was defined in equations (\ref{eq:translated comp law}) and (\ref{eq:geodes for pcomp}), reported here for convenience:

\begin{equation}
\begin{cases}
\frac{d}{dt}\frac{d}{ds}\gamma_{\lambda}^{[p]}(s,t)+\Gamma^{\alpha\beta}\,_{\lambda}(\zeta^{(s)}(t))\frac{d\zeta_{\alpha}^{(s)}(t)}{dt}\frac{d\gamma_{\beta}^{[p]}(s,t)}{ds}=0\\
\gamma_{\lambda}^{[p]}(s,0)=\gamma_{\lambda}^{(k,p)}(s)\\
\gamma_{\lambda}^{[p]}(0,t)=\gamma_{\lambda}^{(q,p)}(t)\\
\gamma_{\lambda}^{[p]}(1,1)=\left(q\oplus_{\ell}^{[p]}k\right)_{\lambda}\end{cases}\label{eq:comp law transl app}\end{equation}

 \begin{equation}
\begin{cases}
\ddot{\zeta}_{\lambda}^{(s)}(t)+\Gamma^{\alpha\beta}\,_{\lambda}(\zeta^{(s)}(t))\dot{\zeta}_{\alpha}^{(s)}(t)\dot{\zeta}_{\beta}^{(s)}(t)=0\\
\zeta_{\lambda}^{(s)}(0)=\gamma_{\lambda}^{(k,p)}(s)\\
\zeta_{\lambda}^{(s)}(1)=\gamma_{\lambda}^{[p]}(s,1)\end{cases}\label{eq:xi definition transl app}\end{equation}

and 

\begin{equation}
\begin{cases}
\frac{d^{2}}{dt^{2}}\gamma_{\lambda}^{(q,p)}(t)+\Gamma^{\mu\nu}{}_{\lambda}(\gamma^{(q,p)}(t))\frac{d\gamma_{\mu}^{(q,p)}(t)}{dt}\frac{d\gamma_{\nu}^{(q,p)}(t)}{dt}=0\\
\gamma^{(q,p)}(0)=p\\
\gamma^{(q,p)}(1)=q\end{cases}\label{eq:geodes for pcomp app}\end{equation}
with a similar formula holding also for $\gamma^{(k,p)}$.

In the following we proceed along the  lines of the previous
section to derive the second order expansion of the translated composition law.

At the zeroth order:
\begin{equation}
\begin{cases}
\frac{d}{dt}\frac{d}{ds}\gamma^{[p]}_{\lambda}(s,t)=0\\
\ddot{\zeta}^{(s)}_{\lambda}(t)=0\end{cases}\end{equation}

Taking into account the boundary conditions these have solution:
\begin{equation}
\begin{cases}
\gamma_{\lambda}^{[p]}(s,t)=\gamma_{\lambda}^{(k,p)} (s)+\gamma_{\lambda}^{q,p}(t)-p_{\lambda}= p_{\lambda}+\left(k_{\lambda}-p_{\lambda}\right)s+\left(q_{\lambda}-p_{\lambda}\right)t \\
\zeta_{\lambda}^{(s)}(t)=p_{\lambda}+\left(k_{\lambda}-p_{\lambda}\right)s+\left(q_{\lambda}-p_{\lambda}\right)t \end{cases}
\end{equation}
So up to the zeroth order the composition law is:
\begin{equation}
\left(q\oplus_{\ell}^{[p]}k\right)_{\lambda}=q_{\lambda}+k_{\lambda}-p_{\lambda}\end{equation}

The first-order expansion of the differential equation in (\ref{eq:comp law transl app}) is 

\begin{equation}
\frac{d}{dt}\frac{d}{ds}\gamma_{\lambda}^{[p]}(s,t)+\tilde\Gamma^{\alpha\beta}\,_{\lambda}\frac{d\zeta_{\alpha}^{(s)}(t)}{dt}\frac{d\gamma_{\beta}^{[p]}(s,t)}{ds}=0
\end{equation}

where $\frac{d\zeta_{\alpha}^{(s)}(t)}{dt}$ and $\frac{d\gamma_{\beta}^{[p]}(s,t)}{ds}$ have to be evaluated at the zeroth order, so that the equation we have to solve is:
\begin{equation}
\frac{d}{dt}\frac{d}{ds}\gamma_{\lambda}^{[p]}(s,t)+\tilde\Gamma^{\alpha\beta}\,_{\lambda}(q_{\alpha}-p_{\alpha})(k_{\beta}-p_{\beta})=0
\end{equation}

Solving for $\gamma_{\lambda}^{[p]}(s,t)$ and taking into account the boundary conditions we find that up to first order:
\begin{equation}
\gamma_{\lambda}^{[p]}(s,t)=\gamma_{\lambda}^{(q,p)}(t)+\gamma_{\lambda}^{(k,p)}(s)-p_{\lambda}-\tilde\Gamma^{\alpha\beta}\,_{\lambda}(q_{\alpha}-p_{\alpha})(k_{\beta}-p_{\beta})st\end{equation}
so that the first-order  composition law takes the form:
\begin{equation}
\left(q\oplus_{\ell}^{[p]}k\right)_{\lambda}=q_{\lambda}+k_{\lambda}-p_{\lambda}-\tilde\Gamma^{\alpha\beta}\,_{\lambda}(q_{\alpha}-p_{\alpha})(k_{\beta}-p_{\beta})\end{equation}

For what concerns the first-order description of the geodesic $\zeta^{(s)}(t)$ we have that it must satisfy the following differential equation:
\begin{equation}
\ddot{\zeta}^{(s)}_{\lambda}+\tilde\Gamma^{\alpha\beta}\,_{\lambda}\dot{\zeta}_{\alpha}^{(s)}\dot{\zeta}_{\beta}^{(s)}=0\end{equation}

which, together with the boundary conditions, has solution:

\begin{equation}
\zeta_{\lambda}^{(s)}(t)=\gamma_{\lambda}^{(k,p)}(s)+\gamma_{\lambda}^{(q,p)}(t)-p_{\lambda}-\tilde\Gamma^{\alpha\beta}\,_{\lambda}(q_{\alpha}-p_{\alpha})(k_{\beta}-p_{\beta})st\end{equation}

The second-order contribution to the differential equation in (\ref{eq:comp law transl app}) is:
\begin{eqnarray}
&&\left(\frac{d}{dt}\frac{d}{ds}\gamma_{\lambda}^{[p]}(s,t)\right)^{(2)}+\tilde\Gamma^{\alpha\beta}\,_{\lambda}\left(\frac{d\zeta_{\alpha}^{(s)}(t)}{dt}\right)^{(0)}\left(\frac{d\gamma_{\beta}^{[p]}(s,t)}{ds}\right)^{(1)}+\tilde\Gamma^{\alpha\beta}\,_{\lambda}\left(\frac{d\zeta_{\alpha}^{(s)}(t)}{dt}\right)^{(1)}\left(\frac{d\gamma_{\beta}^{[p]}(s,t)}{ds}\right)^{(0)}\nonumber \\
&&-\tilde\Gamma^{\alpha\beta}\,_{\lambda}\left(\frac{d\zeta_{\alpha}^{(s)}(t)}{dt}\right)^{(0)}\left(\frac{d\gamma_{\beta}^{[p]}(s,t)}{ds}\right)^{(0)}+\partial^{\theta}\tilde\Gamma^{\alpha\beta}\,_{\lambda}\zeta_{\theta}\left(\frac{d\zeta_{\alpha}^{(s)}(t)}{dt}\right)^{(0)}\left(\frac{d\gamma_{\beta}^{[p]}(s,t)}{ds}\right)^{(0)}=0\end{eqnarray}
where the superscripts refer to the $\ell$ order to be considered for each term.

Substituting the zeroth order results and integrating one finds:
\begin{equation}
\begin{array}{lcl}
\gamma_{\lambda}^{[p]}(s,t)&=&B_{\lambda}(t)+\int ds'A_{\lambda}(s)-\tilde\Gamma^{\alpha\beta}\,_{\lambda}(q_{\alpha}-p_{\alpha})\int dt'\int ds'\left(\frac{d\gamma_{\beta}^{[p]}}{ds}\right)^{(1)}-\tilde\Gamma^{\alpha\beta}\,_{\lambda}\int ds'\int dt'\left(\frac{d\zeta_{\alpha}}{dt}\right)^{(1)}(k_{\beta}-p_{\beta})+\\
&&+\tilde\Gamma^{\alpha\beta}\,_{\lambda}(q_{\alpha}-p_{\alpha})(k_{\beta}-p_{\beta})st-\frac{1}{2}\partial^{\theta}\tilde\Gamma^{\alpha\beta}\,_{\lambda}\left((q_{\theta}-p_{\theta})st^{2}+(k_{\theta}-p_{\theta})s^{2}t+2p_{\theta}s t \right)(q_{\alpha}-p_{\alpha})(k_{\beta}-p_{\beta})\end{array}\end{equation}
Then observing that
\begin{equation}
\begin{cases}
\int ds'\left(\frac{d\gamma_{\beta}}{ds}\right)^{(1)}=\gamma_{\beta}^{(k,p)}(s)-p_{\beta}-\tilde\Gamma^{\gamma\delta}\,_{\beta}(q_{\gamma}-p_{\gamma})(k_{\delta}-p_{\delta})st\\
\int dt'\left(\frac{d\zeta_{\alpha}^{s}}{dt}\right)^{(1)}=\gamma_{\alpha}^{(q,p)}(t)-p_{\alpha}-\tilde\Gamma^{\gamma\delta}\,_{\alpha}(q_{\gamma}-p_{\gamma})(k_{\delta}-p_{\delta})st\end{cases}\end{equation}
and using  the boundary conditions one obtains that up to second order

\begin{equation}
\begin{array}{lcl}
\gamma_{\lambda}(s,t)&=&\gamma_{\lambda}^{(q,p)}(t)+\gamma_{\lambda}^{(k,p)}(s)-p_{\lambda}-\tilde\Gamma^{\alpha\beta}\,_{\lambda}(q_{\alpha}-p_{\alpha})\left(\left(\gamma_{\beta}^{(k,p)}(s)\right)^{(1)}t-p_{\beta}t-\frac{1}{2}\tilde \Gamma^{\gamma\delta}\,_{\beta}\,(q_{\gamma}-p_{\gamma})(k_{\delta}-p_{\delta})s\,t^{2}\right)+\\
&&-\tilde\Gamma^{\alpha\beta}\,_{\lambda}\left(\left(\gamma_{\alpha}^{(q,p)}(t)\right)^{(1)}s-p_{\alpha}s-\frac{1}{2}\tilde\Gamma^{\gamma\delta}\,_{\alpha}(q_{\gamma}-p_{\gamma})(k_{\delta}-p_{\delta})s^{2}t\right)(k_{\beta}-p_{\beta})+\tilde\Gamma^{\alpha\beta}\,_{\lambda}(q_{\alpha}-p_{\alpha})(k_{\beta}-p_{\beta})st\\
&&-\frac{1}{2}\partial^{\theta}\tilde\Gamma^{\alpha\beta}{}_{\lambda}\left(2p_{\theta}s t +(q_{\theta}-p_{\theta})st^{2}+(k_{\theta}-p_{\theta})s^{2}t\right)(q_{\alpha}-p_{\alpha})(k_{\beta}-p_{\beta})\end{array}\end{equation}
So up to second order the translated composition law $q\oplus_{\ell}^{[p]}k$ associated to a connection $\Gamma$
takes the form:
\begin{equation}\begin{array}{lcl}
\left(q\oplus_{\ell}^{[p]}k\right)_{\lambda}&=&q_{\lambda}+k_{\lambda}-p_{\lambda}-\tilde\Gamma^{\alpha\beta}\,_{\lambda}(q_{\alpha}-p_{\alpha})(k_{\beta}-p_{\beta})+\frac{1}{2}\tilde\Gamma^{\alpha\beta}\,_{\lambda}\tilde\Gamma^{\gamma\delta}\,_{\beta}(q_{\alpha}-p_{\alpha})(q_{\gamma}-p_{\gamma})(k_{\delta}-p_{\delta})\\
&&+\frac{1}{2}\tilde\Gamma^{\alpha\beta}\,_{\lambda}\tilde\Gamma^{\gamma\delta}\,_{\alpha}(q_{\gamma}-p_{\gamma})(k_{\delta}-p_{\delta})(k_{\beta}-p_{\beta})-\frac{1}{2}\partial^{\theta}\tilde\Gamma^{\alpha\beta}\,_{\lambda}\left(q_{\theta}+k_{\theta}\right)(q_{\alpha}-p_{\alpha})(k_{\beta}-p_{\beta})\end{array}\end{equation}

Let us now verify that the connection can be derived from the translated
composition law by using:

\begin{equation}
\Gamma^{\mu\nu}{}_{\lambda}(p)=-\frac{\partial}{\partial q_{\mu}}\frac{\partial}{\partial k_{\nu}}\left(q\oplus_{\ell}^{[p]}k\right)_{\lambda}|_{q=k=p}\label{eq:Famigerata pcomp app}\end{equation}

The the derivatives in the above equation give:
\begin{equation}\begin{array}{lcl}
 \frac{\partial}{\partial q_{\mu}}\frac{\partial}{\partial k_{\nu}}\left(q\oplus_{\ell}^{[p]}k\right)_{\lambda}&=&-\tilde\Gamma^{\mu\nu}\,_{\lambda}+\frac{1}{2}\tilde\Gamma^{\alpha\beta}\,_{\lambda}\tilde\Gamma^{\gamma\nu}\,_{\beta}\left(\delta_{\alpha}^{\mu}q_{\gamma}+q_{\alpha}\delta_{\gamma}^{\mu}-\delta_{\alpha}^{\mu}p_{\gamma}-p_{\alpha}\delta_{\gamma}^{\mu}\right)+\frac{1}{2}\tilde\Gamma^{\alpha\beta}\,_{\lambda}\tilde\Gamma^{\mu\delta}\,_{\alpha}\left(\delta_{\delta}^{\nu}k_{\beta}+k_{\delta}\delta_{\beta}^{\nu}-\delta_{\delta}^{\nu}p_{\beta}-p_{\delta}\delta_{\beta}^{\nu}\right)\\
&&-\frac{1}{2}\partial^{\theta}\tilde\Gamma^{\alpha\beta}\,_{\lambda}\left(\left(\delta_{\alpha}^{\mu}q_{\theta}+q_{\alpha}\delta_{\theta}^{\mu}\right)\delta_{\beta}^{\nu}+\left(\delta_{\theta}^{\nu}k_{\beta}+k_{\theta}\delta_{\beta}^{\nu}\right)\delta_{\alpha}^{\mu}-\delta_{\theta}^{\mu}p_{\alpha}\delta_{\beta}^{\nu}-\delta_{\theta}^{\nu}\delta_{\alpha}^{\mu}p_{\beta}\right)\end{array}\end{equation}

so that:

\begin{equation}
-\frac{\partial}{\partial q_{\mu}}\frac{\partial}{\partial k_{\nu}}\left(q\oplus_{\ell}^{[p]}k\right)_{\lambda}|_{q=k=p}=\tilde\Gamma^{\mu\nu}\,_{\lambda}+\partial^{\theta}\tilde\Gamma^{\mu\nu}\,_{\lambda}p_{\theta}=\Gamma^{\mu\nu}\,_{\lambda}(p)\end{equation}

\end{document}